\documentclass[aps,prd,a4paper,reprint,nofootinbib,superscriptaddress,floatfix,preprintnumbers]{revtex4-1}

\usepackage{hyperref}
\usepackage{amsmath}
\usepackage{amsfonts}
\usepackage{amssymb}
\usepackage{color}
\usepackage[utf8]{inputenc}
\usepackage{graphicx}
\usepackage{microtype}
\usepackage{siunitx}
\usepackage{soul}

\begin{document}

\begin{flushleft}
LAPTH-001/20
\end{flushleft}

\title{Cross-correlating galaxy catalogs and gravitational waves: a tomographic approach}
\date{\today}

\author{Francesca Calore}
\affiliation{Univ.~Grenoble Alpes, USMB, CNRS, LAPTh, F-74000 Annecy, France}
\author{Alessandro Cuoco}\email{alessandro.cuoco@unito.it}
\affiliation{Dipartimento di Fisica, Universit\`a di Torino, Via P. Giuria 1, 10125 Torino, Italy}
\affiliation{Istituto Nazionale di Fisica Nucleare, Sezione di Torino, Via P. Giuria 1, 10125 Torino, Italy}
\author{Tania Regimbau}
\affiliation{LAPP, Univ.~Grenoble Alpes, USMB, CNRS/IN2P3, F-74000 Annecy, France}
\author{Surabhi Sachdev}
\affiliation{Department of Physics, The Pennsylvania State University, University Park, PA 16802, USA}
\affiliation{Institute for Gravitation and the Cosmos, The Pennsylvania State University, University Park, PA 16802, USA}
\author{Pasquale Dario Serpico}
\affiliation{Univ.~Grenoble Alpes, USMB, CNRS, LAPTh, F-74000 Annecy, France}
\begin{abstract}
Unveiling the origin of the coalescing binaries detected via gravitational waves (GW) is challenging, notably if no multi-wavelength counterpart is detected. One important  diagnostic tool is the coalescing binary distribution with respect to the large scale structures (LSS) of the universe, which we quantify via 
  the cross-correlation of galaxy catalogs with  GW ones. By using both existing and forthcoming galaxy catalogs 
  and using realistic Monte Carlo simulations of GW events, we find that  the cross-correlation signal  should be marginally detectable in a 10-year data taking of advanced LIGO-Virgo detectors at design sensitivity, at least for binary neutron star mergers. The expected addition of KAGRA and LIGO-India to the GW detector network
would allow for a firmer detection of this signal and, in combination with future cosmological surveys, would also permit the detection of  cross-correlation for coalescing black holes. 
Such a measurement may  unveil, for instance, a primordial origin of coalescing black holes. 
To attain this goal, we find that it is crucial to adopt a tomographic approach 
and to reach a sufficiently accurate localization of GW events.  The depth of forthcoming surveys will be fully exploited by third generation GW detectors such as the Einstein Telescope or the Cosmic Explorer, which will allow one to perform precision studies of the coalescing black hole  LSS distribution and attain rather advanced model discrimination capabilities.

\end{abstract}

\maketitle
\section{Introduction}

New windows in astrophysics usually address existing questions, but also suggest new tools for diagnostic  and raise new puzzles.
This is the case with gravitational waves (GW):  On the one hand, the multi-messenger detection of a neutron star-neutron star merger (BNS)~\cite{GBM:2017lvd} had important implications for a wide range of topics, from $r$-process nucleosynthesis to modified gravity theories. On the other hand, both the progenitors of coalescing binary black holes (BBH) and the environments where they have originated remain mysterious~\cite{LIGOScientific:2018jsj} (see for instance~\cite{Mapelli:2018baq} for a compact review).

At the same time, cosmological applications of this new observational window are routinely discussed.
A notable example of the possible synergy between cosmology and multimessenger astronomy is the cross-correlation between galaxy surveys and 
future GW catalogs~\cite{Namikawa:2015prh,Oguri:2016dgk,Namikawa:2016edr,Raccanelli:2016cud,Scelfo:2018sny,Mukherjee:2018ebj} (or the stochastic GW background~\cite{Cusin:2019jpv,Mukherjee:2019wcg,Canas-Herrera:2019npr,Alonso:2020mva}).
Until now, studies using the cross-correlation have mostly focused on cosmological parameter inference (e.g. non-gaussianities in~\cite{Namikawa:2015prh}, Hubble constant and dark energy in~\cite{Oguri:2016dgk}), or constraints on BBH progenitors (e.g.~\cite{Raccanelli:2016cud,Scelfo:2018sny}). In particular, recent analyses by ~\cite{Raccanelli:2016cud,Scelfo:2018sny} have emphasized that the distribution of BBH with respect to large scale structures (LSS) of the universe, as traced by galaxy catalogs, can disentangle a primordial black hole origin from an astrophysical one. This can be quantified via a dimensionless parameter known as the linear bias $b$, whose definition will be given in Sec.~\ref{method}.
As discussed in~\cite{Raccanelli:2016cud,Scelfo:2018sny}, a bias $b \simeq 1$ is expected from primordial BBH tracing the dark matter LSS, a $b<1$ is typical of low mass halos favored when primordial black holes form binaries only at late times in virialized objects, while a $b>1$ characterizes stellar origin BBH harbored in luminous and massive galaxies, which are biased tracers of LSS. The most recent study by~\cite{Scelfo:2018sny} has found only marginal discrimination capabilities (typical signal to noise ratio of $\sim 1-10$) between two benchmark cases with $b=1$ and $b=1.5$, which improve with future galaxy surveys and GW detectors. The authors have performed a joint cosmological parameters/BBH bias analysis, assessing the impact of including Planck data as well as lensing and relativistic effects, while adopting a parametric approach for the number of GW events and angular resolution of the GW catalog. 

Here we revisit the issue of the determination of the bias, for different classes of binary mergers. We adopt a complementary approach: On the one hand, we fix the cosmology to the Planck best fit, since for a long time to come we expect GW data to provide only modest help to cosmological parameter determination -- at least in minimal cosmological models. This choice also implies that lensing and relativistic effects have tiny impacts  (per-cent level) on the results~\cite{Scelfo:2018sny}, and can be neglected at the price of a (rather small) underestimate of the error. 
On the other hand, instead of using simplified parameterizations for the GW events redshift distribution   and angular reconstruction capabilities, as done in~\cite{Scelfo:2018sny},  we decide to use realistic estimates of these quantities based on the expected performances of GW detectors  obtained from dedicated simulations. 
 Furthermore, we do not limit ourselves to the BBH channel nor to specific pre-defined ranges of redshift $z$ -- as opposed to previous studies that focused either on the cumulative signal at low redshifts~\cite{Namikawa:2016edr}, or on coarse-grained distribution in $z$~\cite{Namikawa:2015prh}, or on high $z$ only ~\cite{Scelfo:2018sny}.

More specifically:
\begin{itemize}
\item[(i)] We realistically simulate catalogs of GW events, taking into account the full star formation history. 
Furthermore, we fully model selection effects for each event, rather than assuming a sharp detectability distance. 

\item[(ii)] Besides BBH, we extend our study to all type of binaries, including also BNS and black hole-neutron star (BHNS). We compare prospects and performances of both existing and forthcoming catalogs. 

\item[(iii)] We highlight the relevance of  the diagnostic power of a binning in $z$ (\emph{tomographic approach}), with a focus on the relatively low-$z$ range.  

\item[(iv)] We do not assume a sharp angular resolution cutoff in multipole ($\ell$) space.
Instead, we adopt a realistic angular resolution derived from  a dedicated Monte Carlo study of the reconstruction of the GW events in a network of detectors. In this respect, we adopt different detector configurations as benchmarks and study the dependence of the results from the various benchmarks.

\end{itemize}

This article is structured as follows. In Sec.~\ref{method} we introduce the formalism. In Sec.~\ref{mockGW} we describe the generated distribution of GW events, as well as the detector configuration benchmarks used. In Sec.~\ref{catalogs} we describe the (existing or forecasted) catalogs used. In Sec.~\ref{results} we present our results, while in Sec.~\ref{conclusions} we draw our conclusions.

\section{Methodology and Formalism}\label{method}

We consider the cross-correlation between the distribution of GW events and  the sky-projected  spatial  distribution of galaxies (or more generally, different types of extragalactic objects).  The cross angular power spectrum (CAPS) between number density fluctuations in a population of discrete objects of two catalogs $a$ and $b$ can be expressed in terms of the multipoles, $C_\ell$:
\begin{equation}
C^{ab}_{\ell} =\frac{2}{\pi}\int k^2 G_{a,\ell}(k)G_{b,\ell}(k){\rm d}k\,,\label{Cell}
\end{equation}
where for both catalogs $i = a$ or $b$
\begin{equation}
G_{i,\ell}(k)=\int{\rm d}z\, \frac{{\rm d}N_i}{{\rm d}z}(z) j_\ell(k\chi(z))\, \delta_i(k,z) \,,\label{Gell}
\end{equation}
where $a$, $b$, are  galaxies and GW events, respectively.
In the expression above, $\chi(z)$ is the comoving distance to redshift $z$, 
${\rm d}N_i/{\rm d}z$ is the redshift  distribution of the objects (normalized so that $\int {\rm d}N_i/{\rm d}z \,  {\rm d}z=1$),
$j_\ell$ the spherical Bessel functions of order $\ell$,
and $\delta_i(k,z)$ is the relative density fluctuation of field $i$ in Fourier space.
The density fluctuation of field $i$ can be related to the
matter density fluctuations $\delta_m$ as  $\delta_i(k,z) = b_i(z)\, \delta_m(k,z)$.
This relation defines  the linear bias $b_i(z)$, which is assumed
to be scale- (i.e., $k$-) independent, but possibly redshift-dependent.
It does, thus, indicate the extent to  which the field $i$ traces the LSS.

Taking the ensemble average  and using the Limber approximation the previous equations can be combined \cite{Fornengo:2013rga} into
\begin{equation}
C^{ab}_{\ell} = \int \frac{{\rm d}\chi}{\chi^2}  P\left(k=\frac{\ell}{\chi},z(\chi)\right)  \prod_{i=a,b} W_i(\chi) b_i(z(\chi)) 
 \label{Cellcomb}
\end{equation}
where 
\begin{equation}
W_i(\chi) \equiv \frac{{\rm d}N_i}{{\rm d}z} \,\frac{ {\rm d}z}{{\rm d}\chi} =  \frac{{\rm d}N_i}{{\rm d}z} \, \frac{ H(z(\chi))}{c}
\end{equation}
is the window function of the field $i$, normalized so that $\int {\rm d}\chi W_i(\chi) =1$,
$H(z)$ is the Hubble function, and $P(k,z)$ is the matter power spectrum as function
of redshift. 
Similarly to the above, we can define the autocorrelation power spectra (APS), $C_\ell^{aa}$, $C_\ell^{bb}$.

The model is fully specified by the ${\rm d}N/{\rm d}z$ (from GW simulations and galaxy catalogs), 
the biases $b_i(z)$ (our unknown and the galaxy catalog one), and the cosmology assumed. The bias of the galaxy catalog, $b_a(z)$, can be typically measured  at a good level of precision (better than 10\%) from the autocorrelation 
of the galaxies in the catalog itself, and we will consider it as a fixed quantify in the following.
The bias of the GW events, $b_b(z)$, is instead   our main unknown, which can be constrained from the
cross-correlation analysis. In the following, for simplicity we will drop the $b$ underscript and  write it simply as $b(z)$.
For the cosmological parameters we use the latest determination from the Planck Collaboration \cite{Aghanim:2018eyx} with a flat cosmology, $\Omega_k=0$ (namely,  
\mbox{$H_0=67.7$ km~s$^{-1}$~Mpc$^{-1}$,} $\Omega_c h^2=0.1193$, $\Omega_b h^2=0.0224$, $n_s=0.966$, $\sigma_8=0.810$).
We use the software \texttt{CLASS} \cite{Blas:2011rf} with these parameters as input to compute the linear matter power spectrum $P_{\rm lin}(k,z)$.
To derive the fully non-linear $P(k,z)$, required in Eq.~\ref{Cellcomb}, we use the halo model prescription \cite{Cooray:2002dia}.
We compare it also with the more common procedure of using \texttt{Halofit} \cite{Takahashi:2012em} and we found small differences of order 10\%.
We stress that even neglecting completely the non-linear corrections to $P(k,z)$ would hardly affect our results. In fact,  we limit the analysis always to a maximum multipole of $\ell$=200, where non-linear effects are subdominant. 
Note that, although some of the analyses reported below are sensitive to larger multipoles (up $\ell$=1000 in some cases), these are unessential
for the determination of the linear bias $b$ of GW events.

Furthermore, given the above limitations,
we decided to neglect general relativistic effects such as weak lensing, which are expected to yield small corrections, notably at the smallest scales not probed by current experiments ~\cite{Raccanelli:2016cud}.  In the future, one may want to include them to account better for parameter correlations when inferring cosmological parameters and their uncertainties~\cite{Oguri:2016dgk,Scelfo:2018sny}.

In general, the detectability of the cross-correlation signal depends on observational aspects, such as the number of sources, the field of view, and the angular resolution of the two catalogs. A simple analytical procedure to estimate these quantities leads to~\cite{Cuoco:2007sh} (see also \cite{Asorey:2012rd,DiDio:2013sea}):
\begin{eqnarray}
\label{eq:cross}
\left(\frac{\delta C_\ell^{ab}}{C_\ell^{ab}}\right)^2=&&\frac{1}{(2\ell+1)\Delta \ell f_{\rm fov}} 
\Bigg[ 1+\frac{C_\ell^{aa}C_\ell^{bb}}{(C_\ell^{ab})^2} \times\nonumber\\
&&\times\left(1+\frac{C_N^{aa}}{{\cal W}_{\ell,a}^2\,C_\ell^{aa}}\right)\left(1+\frac{C_N^{bb}}{{\cal W}_{\ell,b}^2\,C_\ell^{bb}}\right)  \Bigg]\, ,
\end{eqnarray}
where $\Delta \ell$ is the width of bin in $\ell$-space around $\ell$, and  we introduce the fraction of the sky covered by the most restrictive of the two catalogs 
\begin{equation}
\label{eq:fov}
f_{\rm fov}\equiv \frac{\Omega_{\rm fov}}{4\pi}\,,
\end{equation}
and the Poisson noise arising from the discrete number of sources of the catalog, i.e.
\begin{equation}
\label{eq:noise}
C_N^{ii}\equiv \frac{\Omega_{\rm fov}}{N_{i}}\,,
\end{equation}
with $N_i$  the total number of sources within the catalog $i=a,b$ in the considered field of view.
We also introduce the beam (i.e. the point-spread function)  angular window function ${\cal W}_{\ell,i}$, which, 
for a circular beam with a Gaussian profile can be written as
\begin{equation} \label{eq:beam}
{\cal W}_{\ell,i}\equiv \exp\left(-\frac{\sigma_i^2\ell^2}{2}\right)
\end{equation}
where $\sigma_i$ represents the  mean square root deviation of the Gaussian beam 
 for  catalog $i=a$ or $b$.
For galaxy catalogs, $\sigma_a$ is better than arcsec scale and the window function can be set to unity  for the range of $\ell$ we will consider. 
For completeness, the error on the autocorrelation is
\begin{equation}
\left(\frac{\delta C_\ell^{ii}}{C_\ell^{ii}}\right)^2=\frac{2}{(2\ell+1)\Delta \ell f_{\rm fov}} \left(1+\frac{C_N^{ii}}{{\cal W}_{\ell,i}^2\,C_\ell^{ii}}\right)^2 \,.
\end{equation}

A tomographic approach consists in evaluating Eq.~\ref{Cellcomb} in several redshift bins, instead of over the whole redshift range. 
Spectroscopic or sufficiently precise photometric redshifts for each object are provided for most galaxy catalogs.
GW measurements, instead, provide a determination of the luminosity distance of the object  $d_L$, since 
the amplitude of the GW signal scales with the inverse of the luminosity distance. 
The precision with which the luminosity distance can be determined from GW measurements
will be discussed in Sec.~\ref{bayestar}.
Assuming a given cosmology, then $d_L$ can be translated into a $z$ determination.

The GW bias can be derived constructing a $\chi^2$ estimator~\footnote{ Technically, since the error in the denominator of the $\chi^2$ depends on $b$, the $\chi^2$ has a correction term coming from the $b$-dependent normalization factor of the Gaussian distribution. However,  this term varies very slowly in $b$ and thus has a very small effect. }
\begin{equation}
\chi^2(b)=\sum_{j,\ell} \left[\frac{C_{\ell,j}^{ab}(b=1)-C_{\ell,j}^{ab}(b)}{\delta C_{\ell,j}^{ab}(b)}\right]^2\,,
\label{chi2}
\end{equation}
where the sum runs over redshift bins $j$ and the $\ell$-bins.
With the above expression, we can study how well two different  CAPS models, i.e., corresponding
to two different GW biases, can be distinguished. 
In particular, we investigate how a reference model with $b=1$ 
--the first term in the numerator of the $\chi^2$-- can be distinguished from a
model with a generic $b$.

We will use Eq.~\ref{chi2} in a twofold way: First, to determine the significance level at which the two
cases $b=0$ (isotropy) and $b=1$ can be distinguished, interpreting $\sqrt{\chi^2(0)}$ as the significance with which anisotropy can be detected. Secondly, to compute the error with which a bias $b=1$ could be measured. Such an error is indeed provided, at 1$\sigma$ level, by 
the values $b^*$ so that $\chi^2(b^*)=1$.
Clearly, the results depend on the choice of the bias reference value, although this dependence is mild. 
We note, nonetheless, that the most conservative case from a theoretical point of view (and likely correct at least for the BNS case) would be $b>1$, typically associated to binaries coming from the endpoint of stellar evolution~\cite{Scelfo:2018sny}. 
In this respect,  our reference $b=1$ is slightly conservative,
in the sense that a value larger than 1 gives a larger normalization of the CAPS, and thus
a larger signal which would be easier to detect.

Due to the large field of view of GW detectors and for GW surveys over many years as considered here, the sky coverage is quite close to uniform, justifying our all-sky angular analysis. We therefore neglect the effect of slight non-uniform coverage, which is not crucial for our forecast and could be accounted for in the actual analysis knowing the sky-coverage.

\begin{figure}[t]
\includegraphics[angle=0,width=0.45\textwidth]{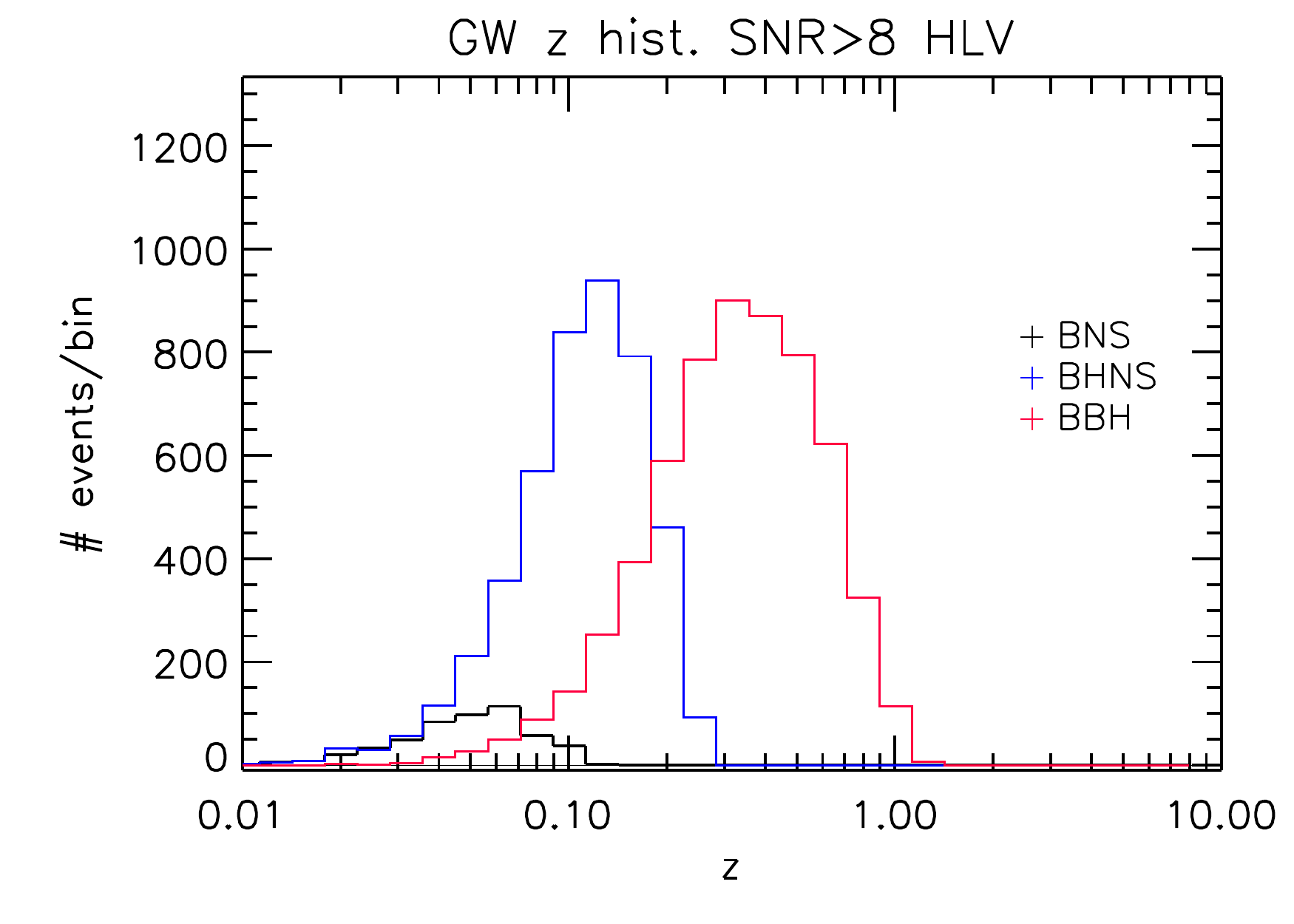}
\includegraphics[angle=0,width=0.45\textwidth]{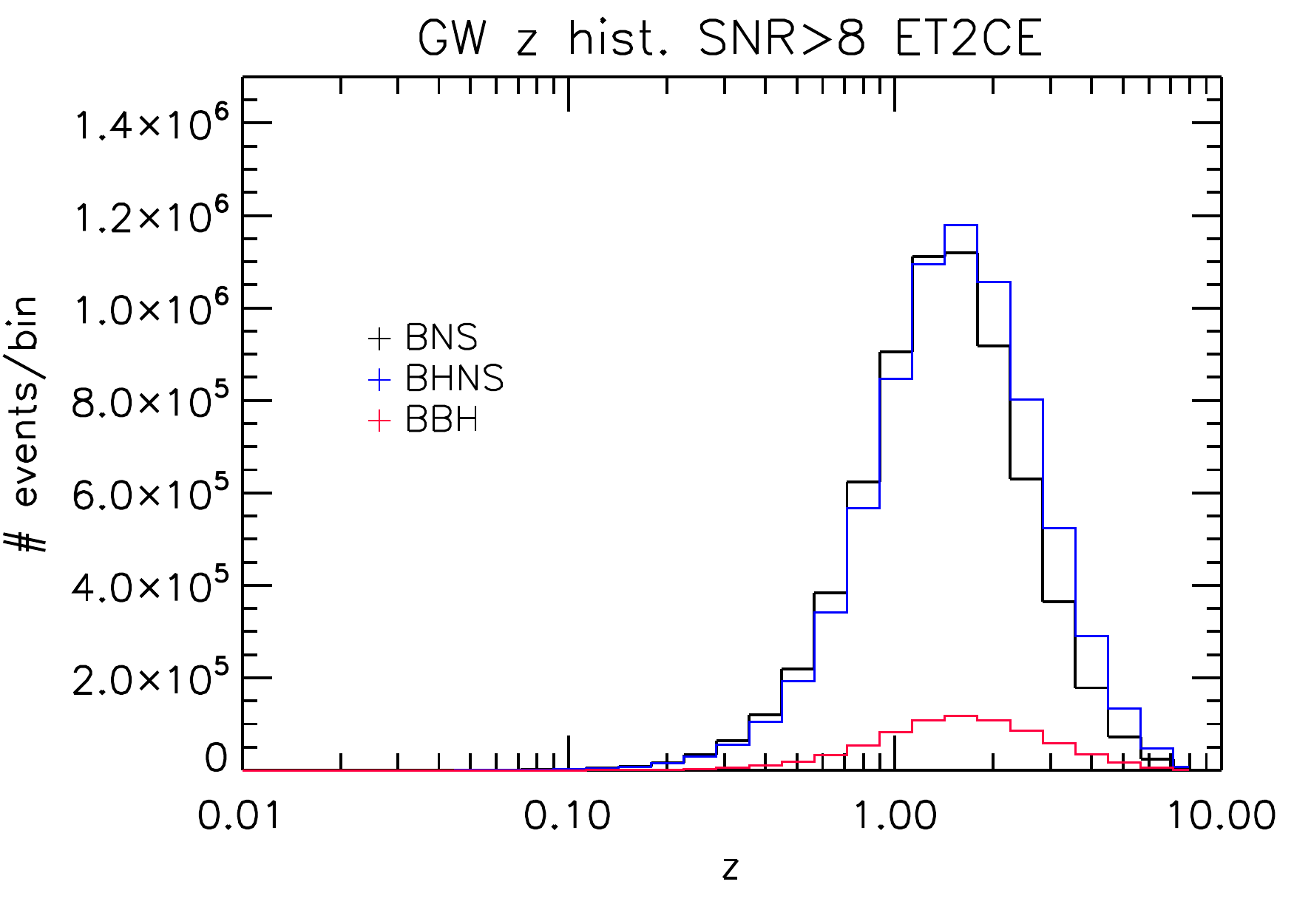}
\caption{Redshift distribution of different classes of GW events for an  SNR cut above 8, for HLV (top) and ET2CE (bottom).} \label{fig:zdistr} 
\end{figure}

\section{Mock gravitational wave catalogs}\label{mockGW}

The GW mock events are generated using a Monte Carlo code developed for the Einstein Mock data challenges~\cite{Regimbau:2012ir,Regimbau:2014uia,Regimbau:2014nxa} and used subsequently for other studies \cite{Meacher:2015iua,Meacher:2015rex,Regimbau:2016ike}.  
We first produce a population of compact binary mergers distributed in the universe, selecting the redshift from the probability distribution function: 
\begin{equation}
p(z)=\frac{R_z(z)}{\int_0^{10} R_z(z) {\rm d}z}\, ,
\end{equation}
constructed by normalizing, in the range $z=[0-10]$, the merger rate per interval of redshift
\begin{equation}\label{red2}
R_z(z) =  \frac{R_m(z)}{1+z} \frac{{\rm d}V(z)}{{\rm d}z} \, .
\end{equation}
The above quantity  is given by the merger rate density $R_m(z)$ times the comoving volume element. The factor $(1+z)$ in the denominator converts the rate in the source frame to the observer frame. 
As in \cite{TheLIGOScientific:2016wyq,Abbott:2017xzg}, we assume that $R_m(z)$ follows the star formation rate of \cite{Vangioni:2014axa} with a time delay $t_d$ whose probability distribution is of the form of $P(t_d) \sim t_d^{-1}$, with a minimum delay time $t_{\min}$ ranging from 20 Myr for BNS to 50 Myr for BBH:
\begin{equation}\label{RnzRf}
R_m(z) \propto \int_{t_{\min}}^{t_{\max}} R_{z_f}(z_f)P(t_d) dt_d\, \mathrm{\,with\,} R_m(0) \equiv R_0 \,,
\end{equation}
where $z_f$ is the redshift of formation of the binary.
The local rate, $R_0$, corresponds to the rates inferred from the detections in the first two LIGO/Virgo observation runs~\cite{LIGOScientific:2018mvr}. We use median rates obtained from the \texttt{GstLAL} pipeline, assuming a Gaussian distribution of the component masses for BNS ($R_0=920 \, {\rm Gpc}^{-3}\, {\rm yr}^{-1}$) and a power-law distribution for BBH ($R_0=56 \, {\rm Gpc}^{-3}\, {\rm yr}^{-1}$). 
For the case of BHNS we consider the upper limit derived assuming fixed component masses of 1.4 M$_\odot$ and 10 M$_\odot$ ($R_0= 600\, {\rm Gpc}^{-3}\, {\rm yr}^{-1}$)~\cite{LIGOScientific:2018mvr}.  Note that a future refined determination of the event rates would translate in a renormalization of the running time required to attain the performances reported in the following. 

After the redshift is selected, we draw the masses of the components from the distributions above, as well as the inclination of the orbit, the polarization and the phase at the merger from uniform distributions~\cite{Regimbau:2012ir}.

In order to determine whether a source can be detected, for a specific observational scenario (see below), we calculate the combined match-filtered signal-to-noise ratio, SNR, as the quadrature sum SNR$^2= \sum_i {\rm SNR}_i^2$ of the individual signal-to-noise ratio in each detector, ${\rm SNR}_i$, cf.~Eq.~18 in~\cite{Regimbau:2012ir}.  
In this paper we assume a 10-year data taking with a duty cycle of 80\% for each detector and we set the detection threshold to ${\rm SNR}>8$, since a modest contamination with noise is tolerable for this specific statistical analysis. Spurious events due to noise are indeed  expected to be uncorrelated with extragalactic structures and would not bias the current forecast.

We consider three possible  configurations for the GW detector network (see also Tab.~\ref{tab:rates}):
\begin{itemize}
\item[I.]  {\bf HLV}: Advanced LIGO-Virgo with three detectors of second generation (2G) at design sensitivity, at the current Hanford, Livingston (LIGO) and Cascina (Virgo) locations. 
Typical samples of detected mock GW events for an SNR above 8 (based on the simulation described above) are $\sim$500 BNS events, 4000 BBH events, and up to 2000 BHNS mergers.
The redshift distribution of the events from the simulation is reported in Fig.~\ref{fig:zdistr} (upper panel).
We see that in this case the BNS events are all cosmologically very close (within $z \lesssim 0.15$), while BBH extend to higher $z$,
up to a maximum of $z\sim 1$ with a peak at $z\sim0.3$. The BHNS redshift distribution is intermediate between the two other classes. 
\item[II.] {\bf HLVIK}: Five 2G detectors, three as above, adding LIGO-India~\cite{Unnikrishnan:2013qwa} and KAGRA~\cite{Somiya:2011np}.
The redshift distribution of detectable sources is very similar to the previous case, extending to slightly higher $z$. Instead, the statistics increases by  roughly 5 times for each class of mergers, i.e. 2500 BNS events and $2 \times 10^{4}$ BBH (and a similar upper limit to the BHNS events).
\item[III.] {\bf ET2CE}: Three detectors of third generation (3G,~\cite{Kalogera:2019sui}), namely one Einstein Telescope (ET,~\cite{Punturo:2010zz,Maggiore:2019uih}) at the Virgo site and two Cosmic Explorers (CE, \cite{Reitze:2019iox}) at the two LIGO sites.
This configuration would allow one to explore redshifts up to $z \sim 5$, with a distribution typically peaking around $z \sim 1$ for all classes.
The number of events will also feature a large increase up to two orders of magnitude with respect to HLVIK, with $\sim 7 \times 10^{6}$ BNS and $\sim 7 \times 10^{5}$ BBH. However, since HLVIK has  already almost full efficiency in detecting BBH below $z\sim0.3$, HLVIK and ET2CE will detect roughly the same number of BBH ($\sim 6000$ in ten years) below $z\simeq0.3$.

\end{itemize}

\begin{table}[t]
\caption{Number of events with SNR$>8$ for a 10-yr data taking period (at 80\% duty cycle) of the three  reference networks of GW detectors, for the three classes of events. Numbers for BHNS are upper limits.}
\label{tab:rates}
\centering
{
\footnotesize
\begin{tabular}{c|c|c|c}
\hline
Case & BNS & BBH & BHNS \\
\hline \hline
HLV & $5\times 10^2$     & $4\times 10^3$  & $<2\times 10^3$  \\
HLVIK & $2.5\times 10^3$     & $2\times 10^4$ & $<2\times 10^4$   \\
ET2CE & $7\times 10^6$     & $7\times 10^5$ & $<5\times 10^6$   \\
\hline
\end{tabular}
}
\end{table}

\begin{figure}[t]
\includegraphics[angle=0,width=0.45\textwidth]{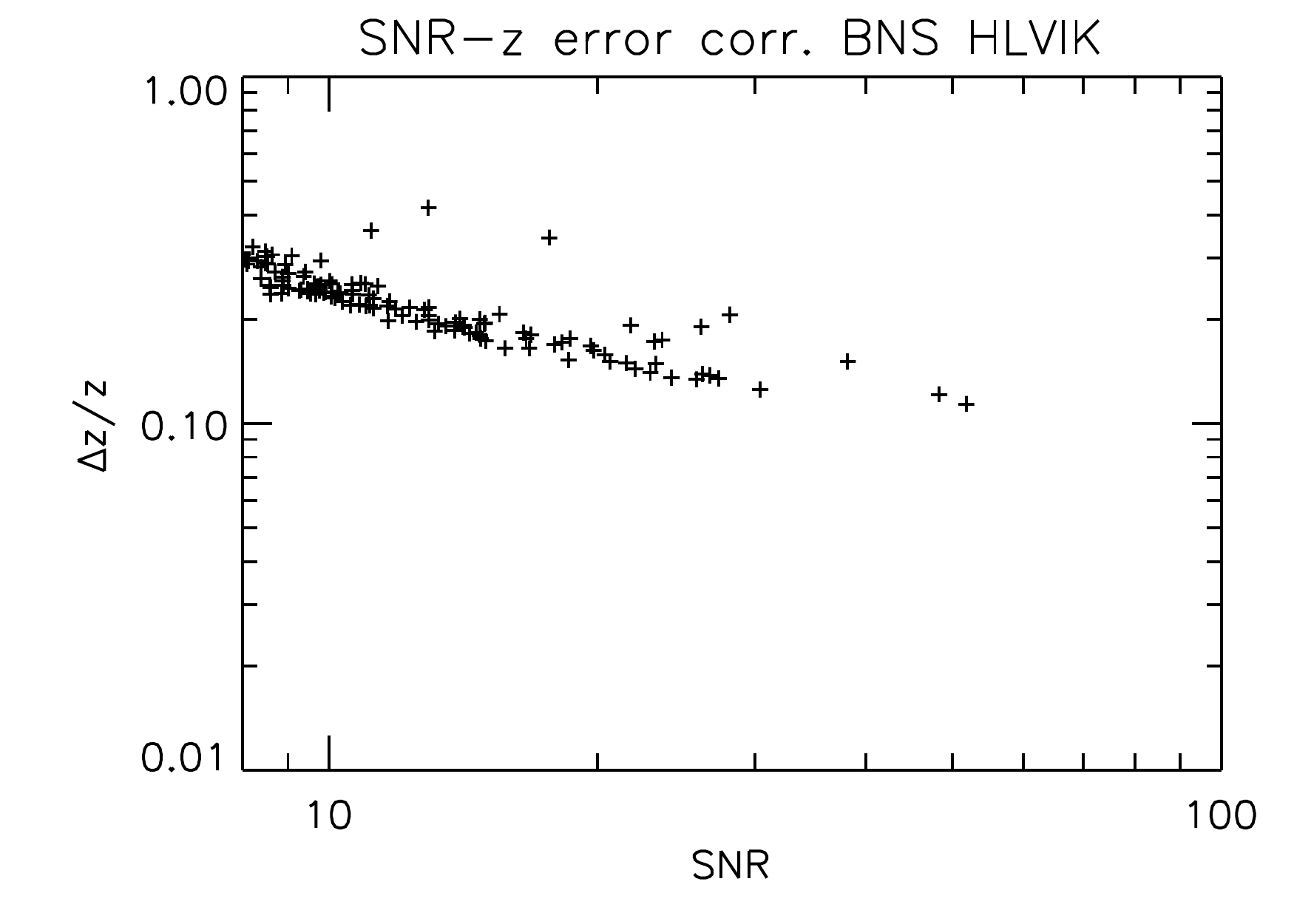}
\includegraphics[angle=0,width=0.45\textwidth]{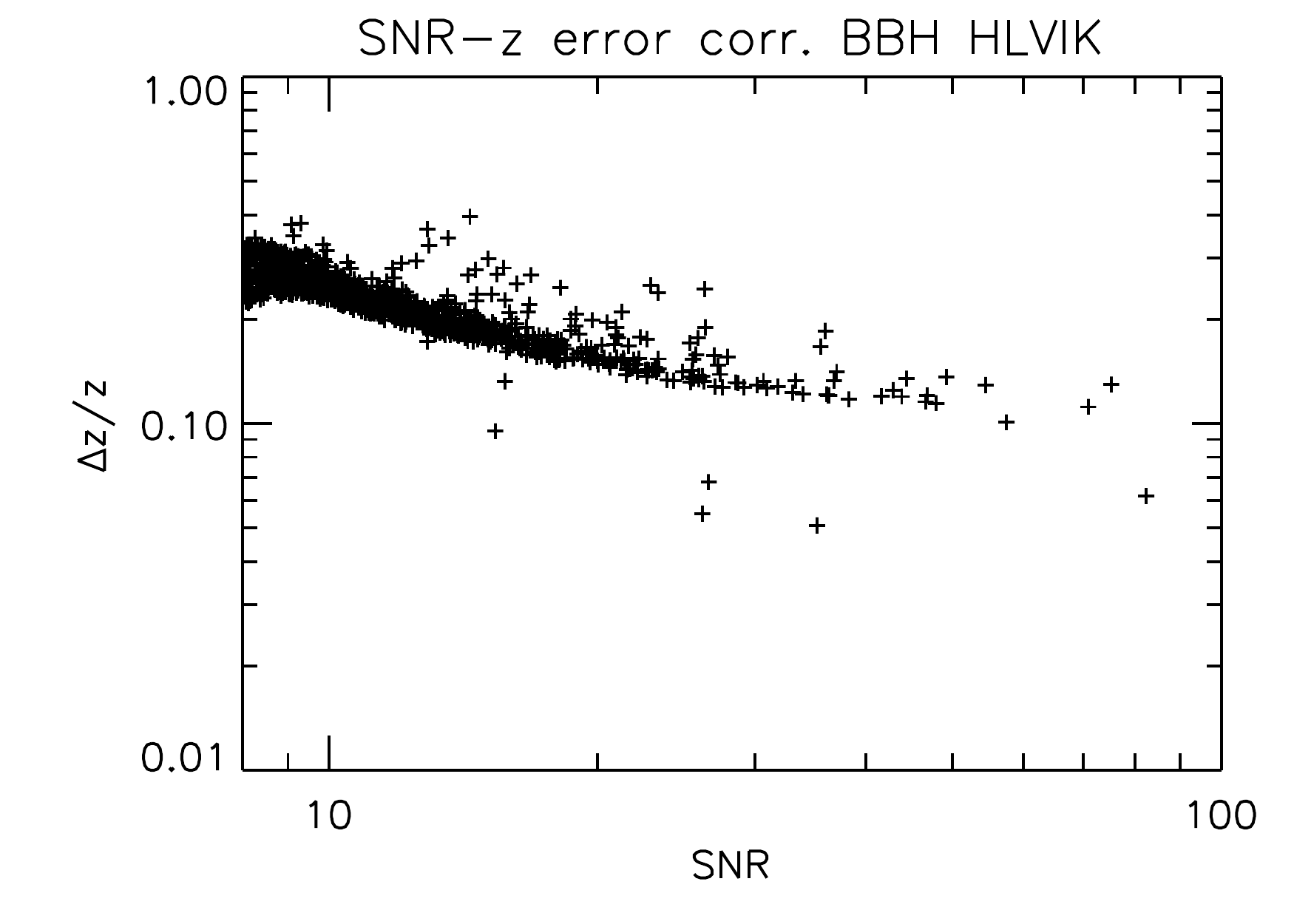}
\includegraphics[angle=0,width=0.45\textwidth]{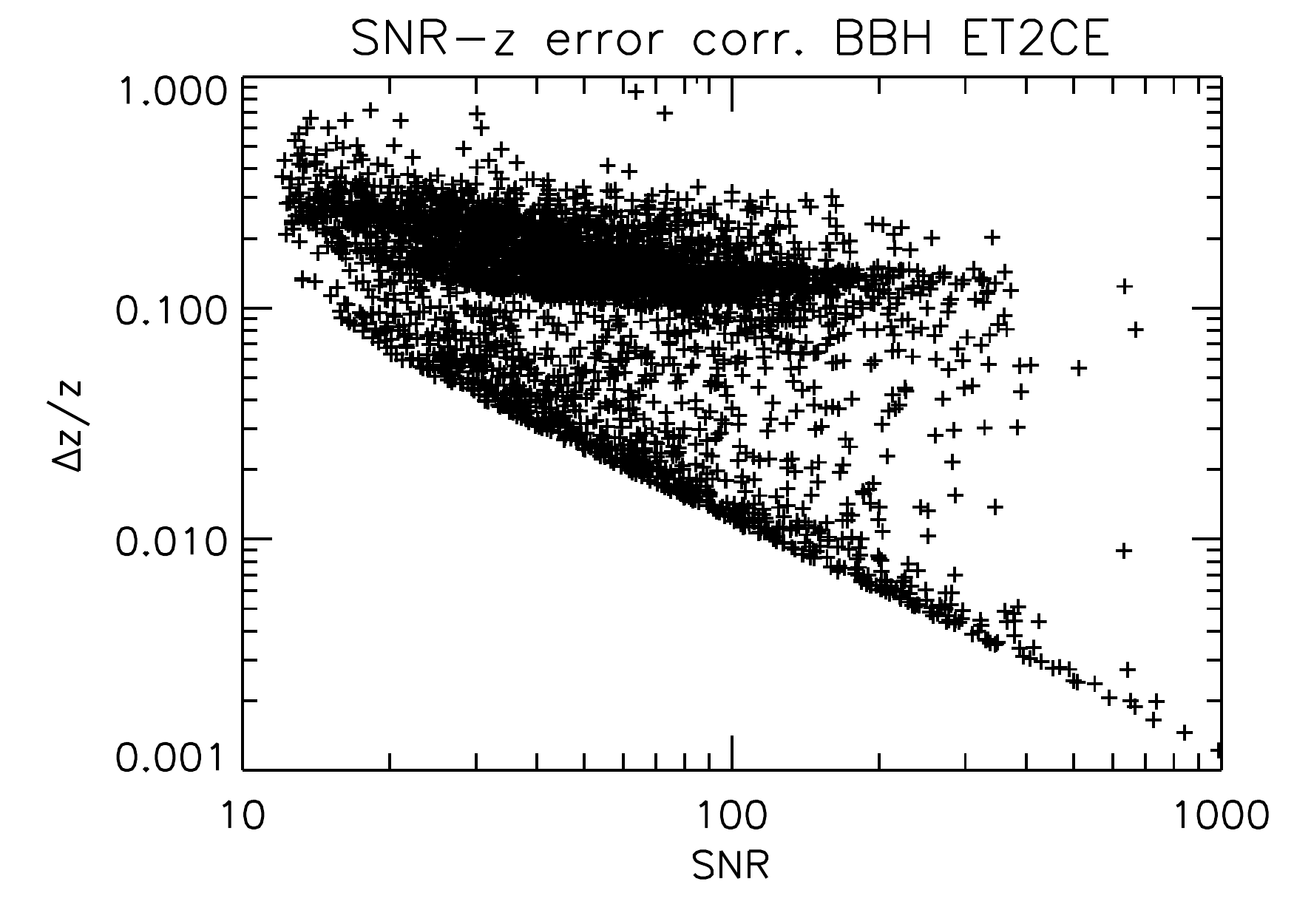}
\caption{Correlation between the relative error on $z$ and SNR for a sample of sources
simulated with \texttt{BAYESTAR}:
about 300 events of the BNS class in the HLVIK case (top panel),
about $10^3$ events of the BBH class in the HLVIK case (middle panel),
and about $1.4 \times 10^4$ events of the  BBH class in case ET2CE (bottom panel).
\label{fig:SNRzerrcorr} }
\end{figure}

\begin{figure}[t]
\includegraphics[angle=0,width=0.45\textwidth]{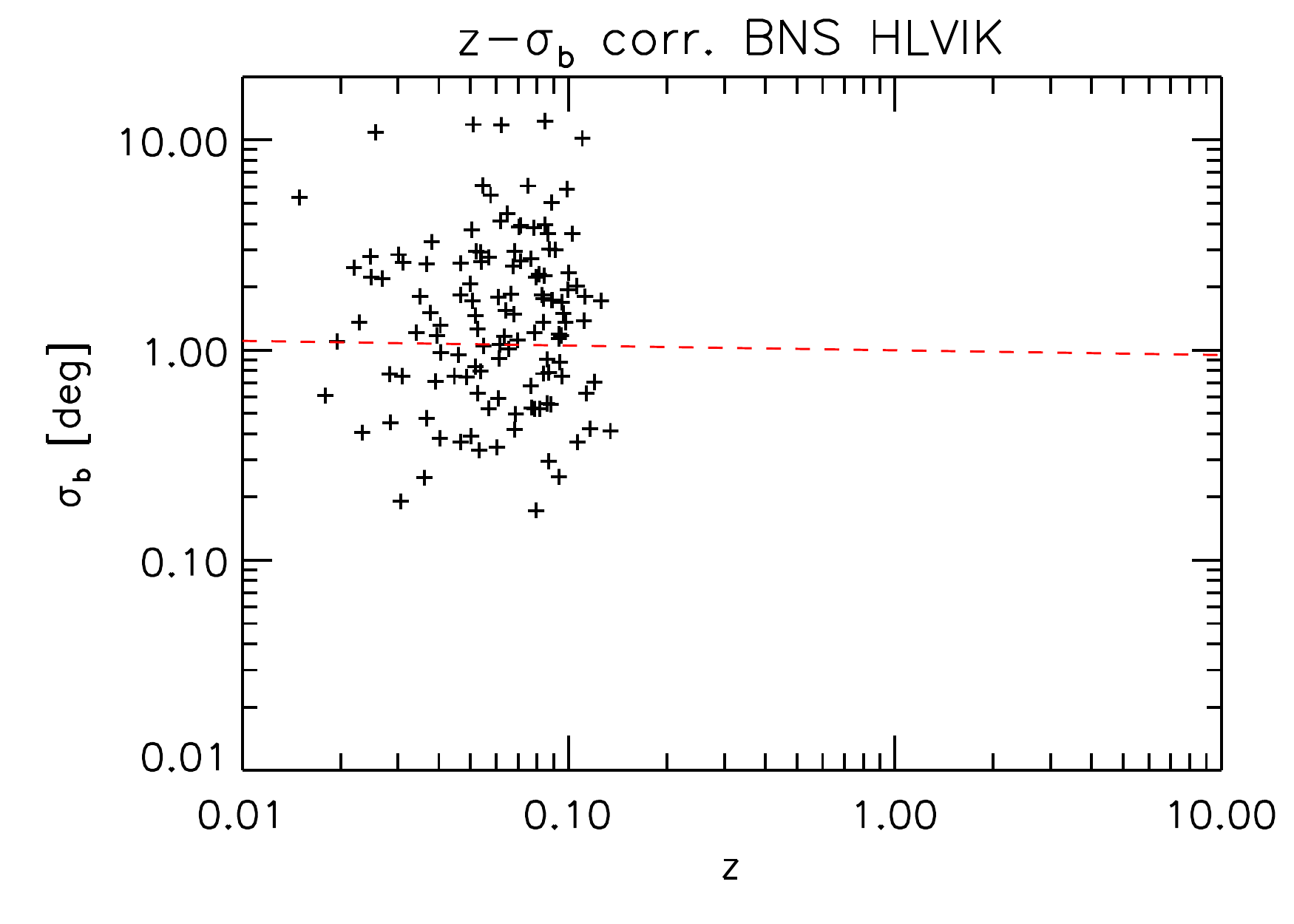}
\includegraphics[angle=0,width=0.45\textwidth]{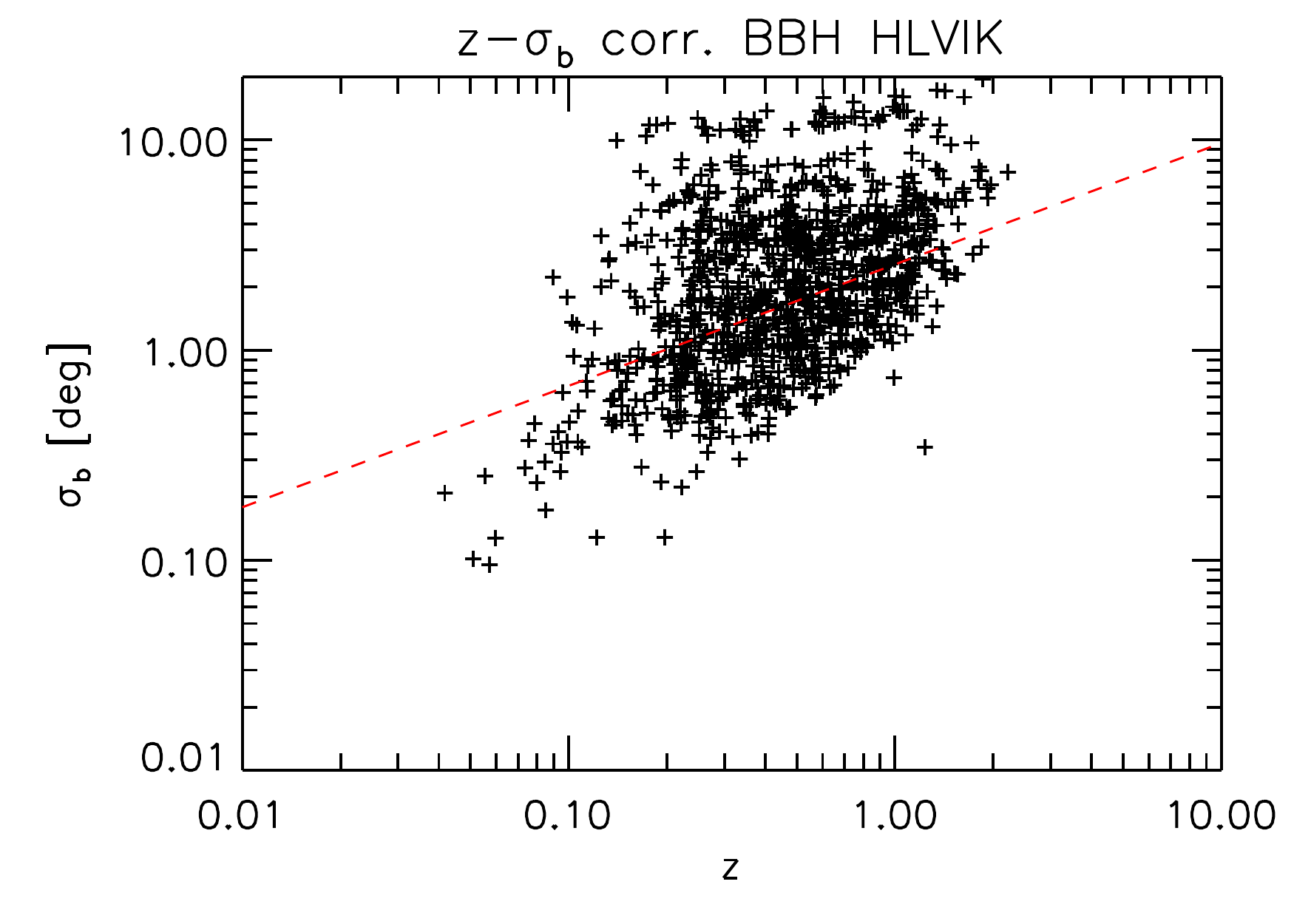}
\includegraphics[angle=0,width=0.45\textwidth]{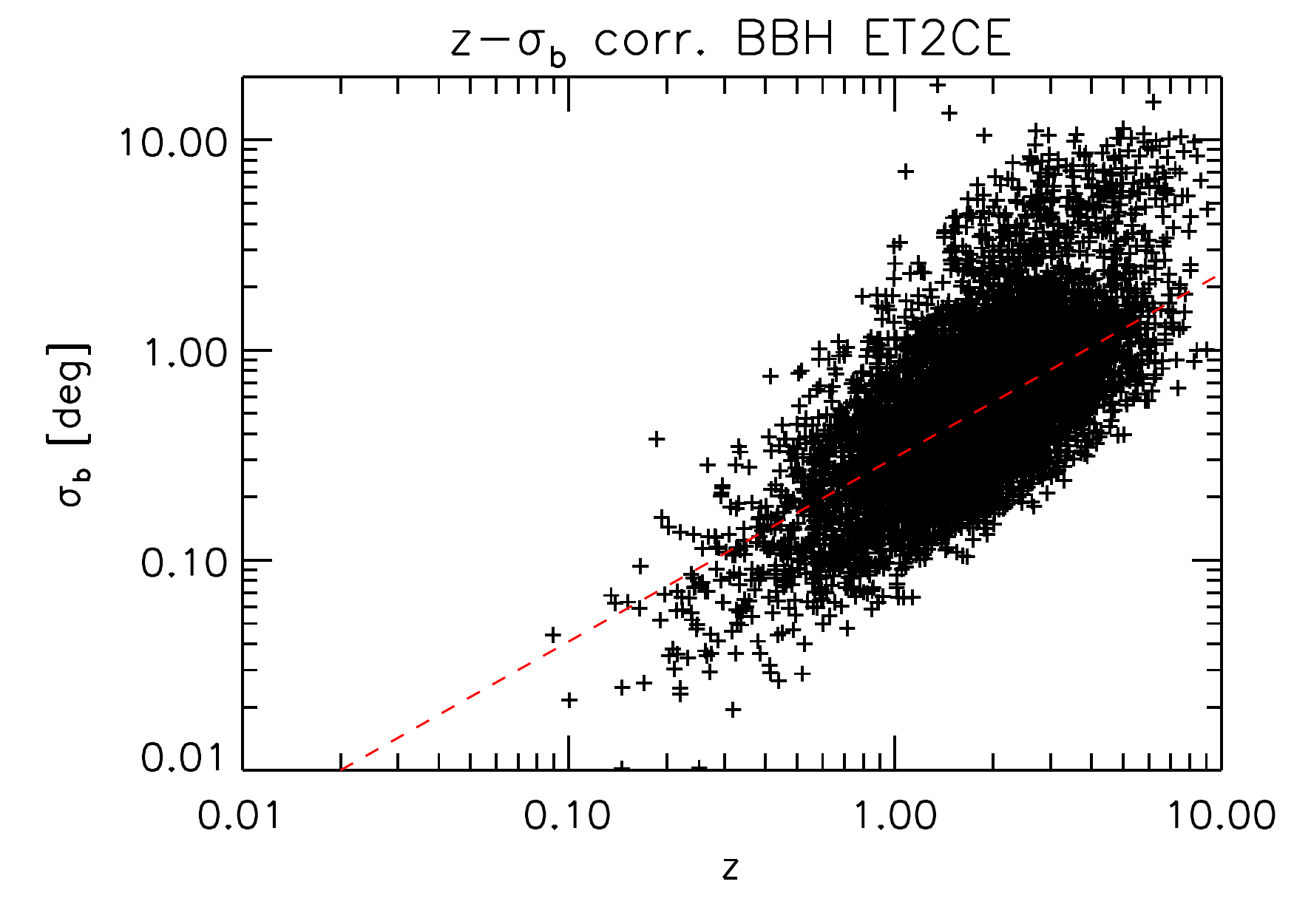}
\caption{Correlation between $\sigma_b$ and $z$ for the BNS and BBH class in the HLVIK case (top and middle panel, respectively), as well as for the  BBH class in the ET2CE case (bottom panel),
for the same sample of events considered in Fig.~\ref{fig:SNRzerrcorr}.
The red line shows the adopted regression fit. 
\label{fig:zsigmacorr}
 }
\end{figure}

\subsection{Positional reconstruction with \texttt{BAYESTAR}}
\label{bayestar}

\begin{figure}[t]
\includegraphics[angle=0,width=0.45\textwidth]{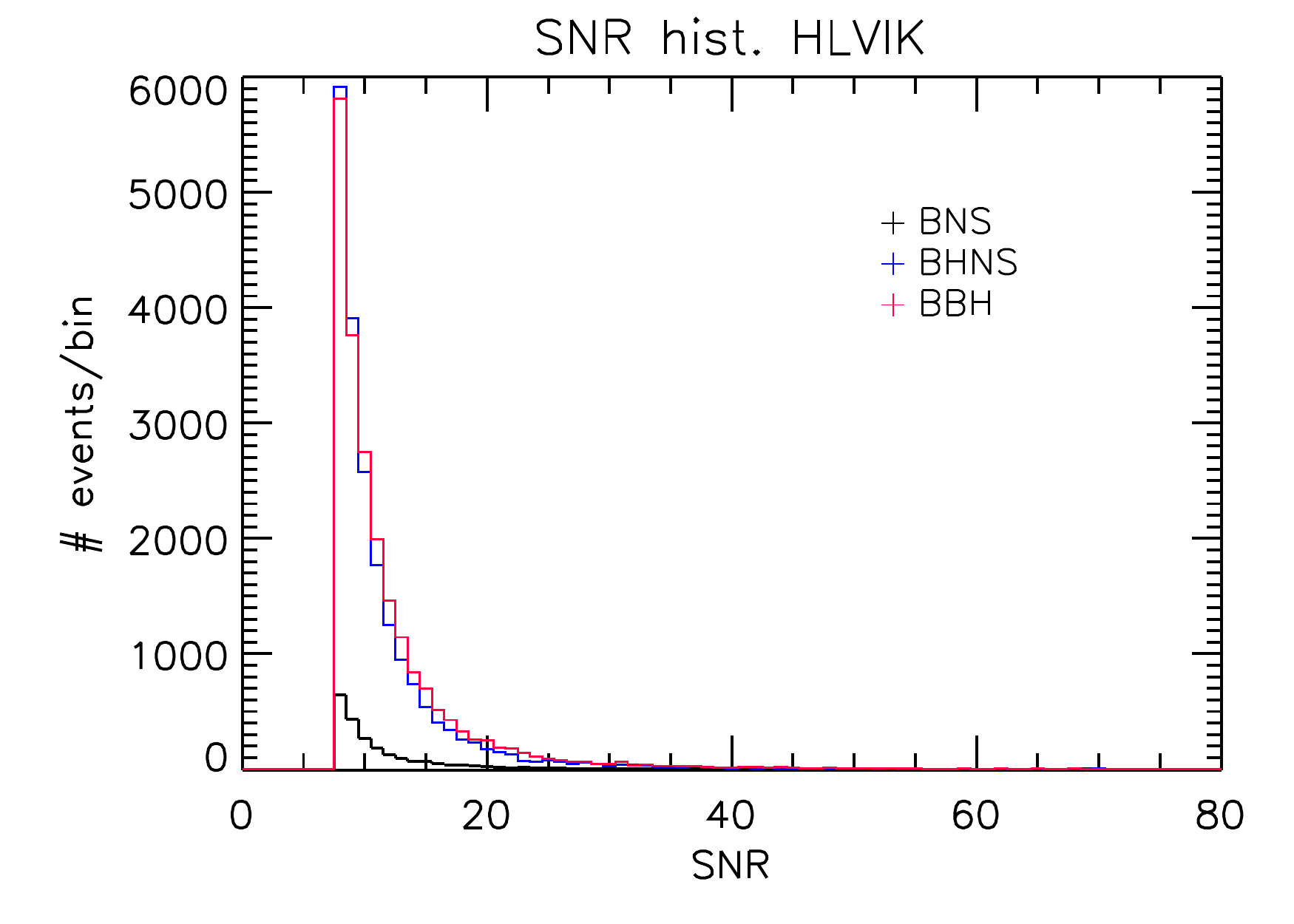}
\includegraphics[angle=0,width=0.45\textwidth]{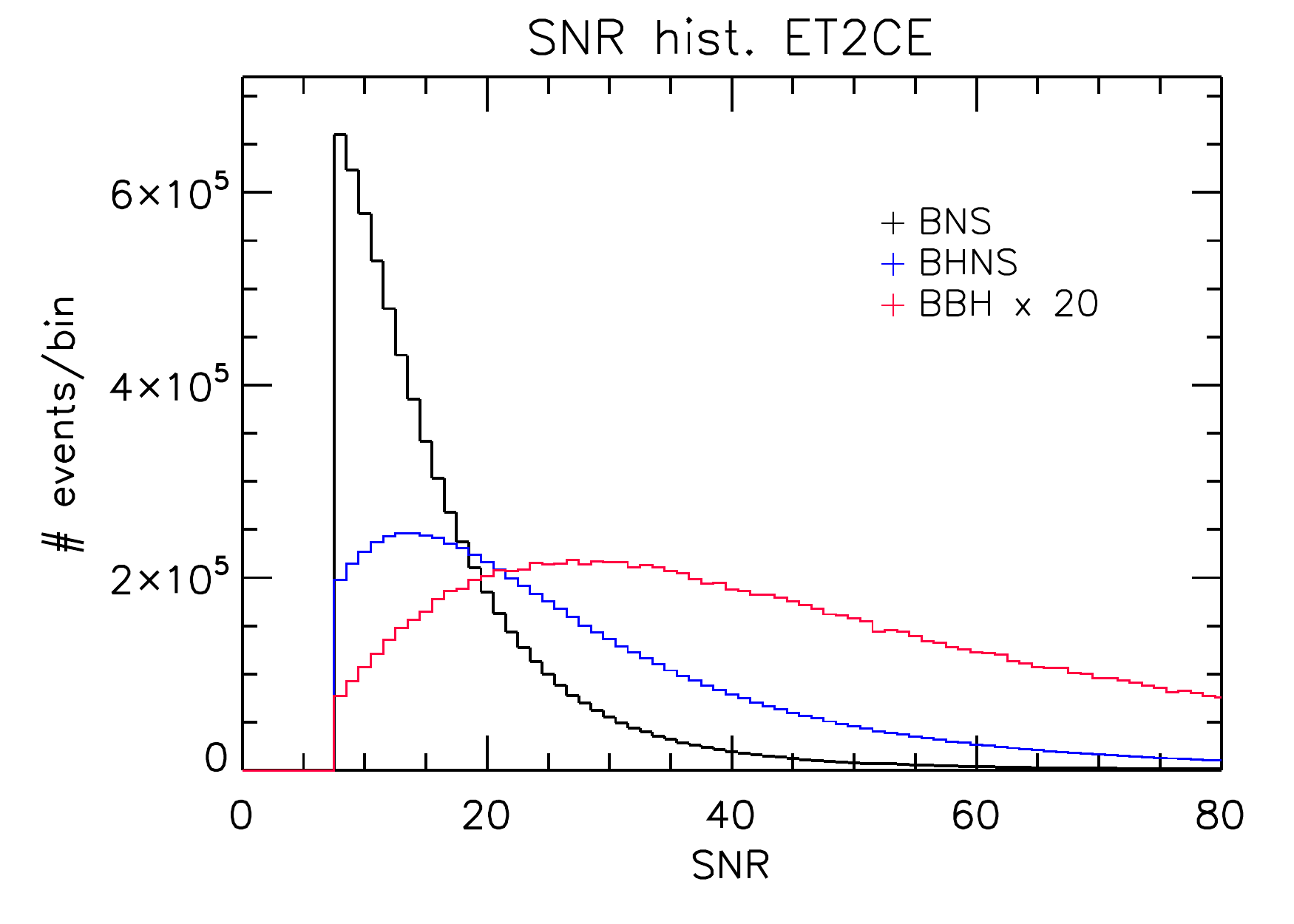}
\caption{SNR distribution for different classes of events for the HLVIK case (top), and the  ET2CE case (bottom),
for the full sample of events in our 10-year data taking simulation.
\label{fig:SNRdist}}
\end{figure}

In order to perform the cross-correlation analysis, it is important to know if using a Gaussian beam is a good approximation, cf.~Eq.~\ref{eq:beam}, and also have an estimate for its width. This is investigated by running  \texttt{BAYESTAR}~\cite{Singer:2015ema}, a fast Bayesian sky localization algorithm that produces probability sky maps that are comparable in accuracy to the computationally expensive Markov chain Monte Carlo methods, like, e.g., \texttt{LALinference}~\cite{Veitch:2014wba}, for full GW parameter estimation. \texttt{BAYESTAR} returns also posterior probabilities for the reconstructed luminosity distance. 

We perform the reconstruction of about 1000 BBH events and 300 BNS for the HLVIK case, $\sim$$1.4 \times 10^{4}$ BBH events and $\sim$ 5000 BNS events for the ET2CE case, 
which is enough to perform statistical studies of the average expected sky localization and redshift error.
For the BHNS case, which is anyway the less reliable and constrained of the cases shown, we just assume for all cases similar performances as for the BBH case, given the expected similarity of their SNR distribution shapes, c.f.~Fig.~\ref{fig:SNRdist}.

In the literature, the error on the luminosity distance and redshift is typically estimated to be inversely proportional to the SNR, for example, $\delta z/z \simeq \delta d_L / d_L \simeq 3/$SNR in Eq.~3 of~\cite{Pozzo:2014qwa}. This corresponds to $\sim 35\%$ for an SNR of 8. 
Such an approximation is only partly confirmed by the \texttt{BAYESTAR} sky localization results, as shown in Fig.~\ref{fig:SNRzerrcorr}:  On the one hand, we observe the predicted 20-30\% error on the reconstructed
redshift (assuming $\Lambda$CDM cosmology); on the other hand, the relative error is weakly dependent on the SNR.
Only for the BBH ET2CE case a second mode is present, with a fraction of the events
showing the expected behaviour  scaling as $\propto$1/SNR.
  This poses a lower limit  on the minimum width of the bins in which the sample can be divided to perform tomography. 
In the following, nonetheless,  we will use quite large bins in $z$, thus below the above mentioned limit.

As for the angular reconstruction, discussions can be found in 
Refs.~\cite{Vitale:2016avz,Pozzo:2014qwa,Vitale:2011zx,Fairhurst:2017mvj,Sidery:2013zua,Schutz:2011tw, Klimenko:2011hz}.
However,  most references  provide only the average containment area of the event at a given level
of significance, typically 68\% or 90\% C.L., but not the full shape of the ``beam", which is in principle required for the full angular analysis.

With our simulations, we are able to study the shape of the full posterior probability distribution of the reconstructed localization.  We find that when the event is reasonably well localized (i.e better than $\sim$ 100 deg$^2$ at 50\% C.L.)
the posterior is very well approximated by a two-dimensional Gaussian.
Furthermore,  in the case HLVIK, with five detectors, the posterior is also circular to a good approximation.
It is thus possible to use the formula in Eq.~\ref{eq:beam}, for a circular beam with Gaussian profile with a width $\sigma_b$.  
For the case ET2CE, with three detectors, instead, the 2D Gaussian remains significantly elongated
and the beam has an elliptical cross-section, rather than circular. 
While the formalism from Sec.~\ref{method} can be extended to include an elliptical beam, this would be an unnecessary
complication for the forecast purpose of this work and we consider the GW beam to be circular also in the ET2CE case. In order to derive $\sigma_b$, we start from the 50\% C.L. localization area, as provided by \texttt{BAYESTAR}, and we convert it to $\sigma_b$ assuming a circular Gaussian beam.

Fig.~\ref{fig:zsigmacorr}  shows that for the BBH cases there is a strong correlation between the redshift $z$ and $\sigma_b$. 
For the BBH HLVIK  case,  we obtain $\sigma_b\simeq 2.56^\circ\,z^{0.58}$ (valid at $z\gtrsim 0.1$) as regression fit of the  points. This rather good angular resolution that we adopt
in the following requires only a mild quality cut, excluding the small fraction (about  ${\cal O}$(15\%)) of events with $\sigma_b >5^\circ$.  We note that this procedure mimics what can be done on real data, where the reconstructed angular resolution is known event-by-event, and it is thus possible to apply angular reconstruction quality cuts on samples of events.
For the BNS HLVIK  case, the angular resolution is almost $z-$independent at the shallow distances accessible to the detectors. If we exclude  BNS events with $\sigma_b >3^\circ$ (constituting about 20\% of the sample), our fit  yields $\sigma_b\simeq 1.0^\circ\,z^{-0.022}$ at $z\gtrsim 0.01$.
The BBH ET2CE case has a very good angular resolution for $z<1$.
This is mainly due to the fact that low-$z$ events have a very large SNR in ET2EC and the angular
resolution is a decreasing function of SNR. 
In deriving the $z$-$\sigma_b$ relation we have excluded the events with $\sigma_b >3^\circ$,
which, as can be seen from Fig.~\ref{fig:zsigmacorr}, constitute a second sub-dominant mode containing only ${\cal O}$(1\%)
of the total events. For $z\gtrsim 0.1$, we obtain $\sigma_b\simeq 0.31^\circ\,z^{0.88}$.
The localization for BNS with ET2CE is typically worse than for BBH. 
In particular, the distribution is bimodal, similarly to the BBH case, but the fraction of events badly reconstructed (with $\sigma_b$ as high as 10$^\circ$) is larger than 50\%.
We attribute the different reconstruction quality for BBH and BNS largely to the
different SNR distributions of the two samples. This is illustrated in the bottom panel of Fig.~\ref{fig:SNRdist},
which shows that the SNR distribution of BNS ET2CE  is peaked at low SNR, qualitatively similar to the event distribution of the HLVIK reported in the top panel of Fig.~\ref{fig:SNRdist}, while the one of BBH
is peaked at SNR $\sim$ 30. 
In other words, the BBH sample seen by ET2EC is virtually complete, with the improved sensitivity resulting in stronger
signals, rather than many more events at low SNR.
Thus, the second mode, which for both BBH and BNS contains events with low SNR, is scarcely populated
for BBH while containing about $\sim 50\%$ of the events in the BNS case. Therefore, apart for improved statistics, we do not expect the ET2EC sample of BNS to show qualitatively new features with respect to HLVIK. For BBH the situation is very different, hence, in what follows, we will focus on the novel possibilities to study this class of events brought by 3G detectors.

 In Tab.~\ref{tab:sigmab}, we quote the parameters of the $\sigma_b(z)$ linear regression fit (in $\log_{10}$ space) for the two classes of events (BNS, BBH) and the two detector configurations HLVIK and ET2CE, obtained considering only the mode with good angular resolution for each case (See Fig.~\ref{fig:zsigmacorr}).
\begin{table}[t]
\caption{Parameters of the $z$-$\sigma_b(z)$ relation, in the form $\log_{10} (\sigma_b(z)) = A + B \, \log_{10}(z)$, with $A \equiv  \log_{10} (\sigma_b(z=1))$ corresponding to the best-fit parameters of the red lines in Fig.~\ref{fig:zsigmacorr}.}
\label{tab:sigmab}
\centering
{
\footnotesize
\begin{tabular}{c|cc|cc}
\hline
Case &BNS & & BBH &\\
\hline
Parameter & A & B & A & B \\
\hline \hline
HLVIK &     -0.00076          &  -0.0224     &        0.4079          &  0.5778  \\
ET2CE &    -0.0317        &  0.5249     &       -0.5109          &  0.8760 \\
\hline
\end{tabular}
}
\end{table}

For the HLV case, the presence of only three detectors combined with a reduced sensitivity translates into an SNR strongly peaked at low values, and more pronounced deviations from a Gaussian reconstructed region are present. For this case, it is also computationally more challenging to collect a statistics of acceptable quality events sufficient for being post-processed via \texttt{BAYESTAR}. Since the angular resolution is mostly influenced by the number and location of the detectors, we used the ET2CE performance 
to estimate an {\it optimistic} proxy for the HLV case, consisting in a  circular equivalent $\sigma_b$ of about 4$^\circ$ and no redshift dependence. Since we will see that the perspectives for a detection in the HLV case are very poor, a more conservative choice would not  alter significantly our conclusions.

Note that, for BNS (and possibly for BHNS), electromagnetic counterparts will be available for a certain
fraction of the events, which will depend crucially on the multi-messenger efforts of the astronomical community
and the reliability of current models. This fraction is difficult to estimate at the moment and will be thus conservatively neglected in the following. It is clear, though, that for these events a very precise redshift and angular determination will
be available, allowing one to use this sub-sample  for more precise angular and  tomographic analyses.

 \begin{figure}[t]
\begin{center}
\begin{tabular}{c}
\includegraphics[angle=0,width=0.45\textwidth]{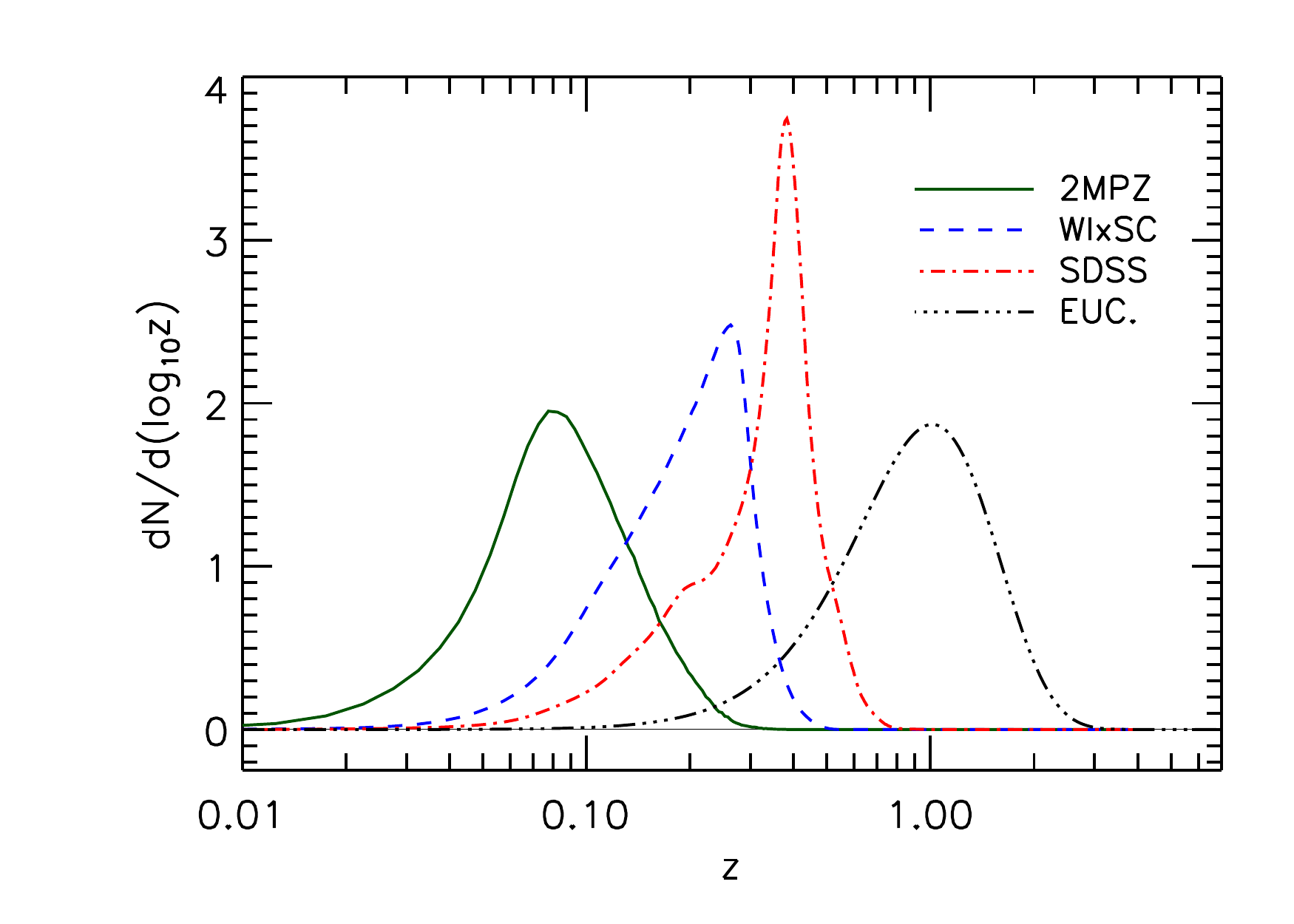}
\end{tabular}
\end{center}
\caption{Redshift distribution, ${\rm d}N/{\rm d}z$, of the galaxy catalogs used for the cross-correlation. All distributions are normalized to 1.  
\label{fig:catal}}
\end{figure}

 \begin{figure}[t]
\begin{center}
\begin{tabular}{c}
\includegraphics[angle=0,width=0.48\textwidth]{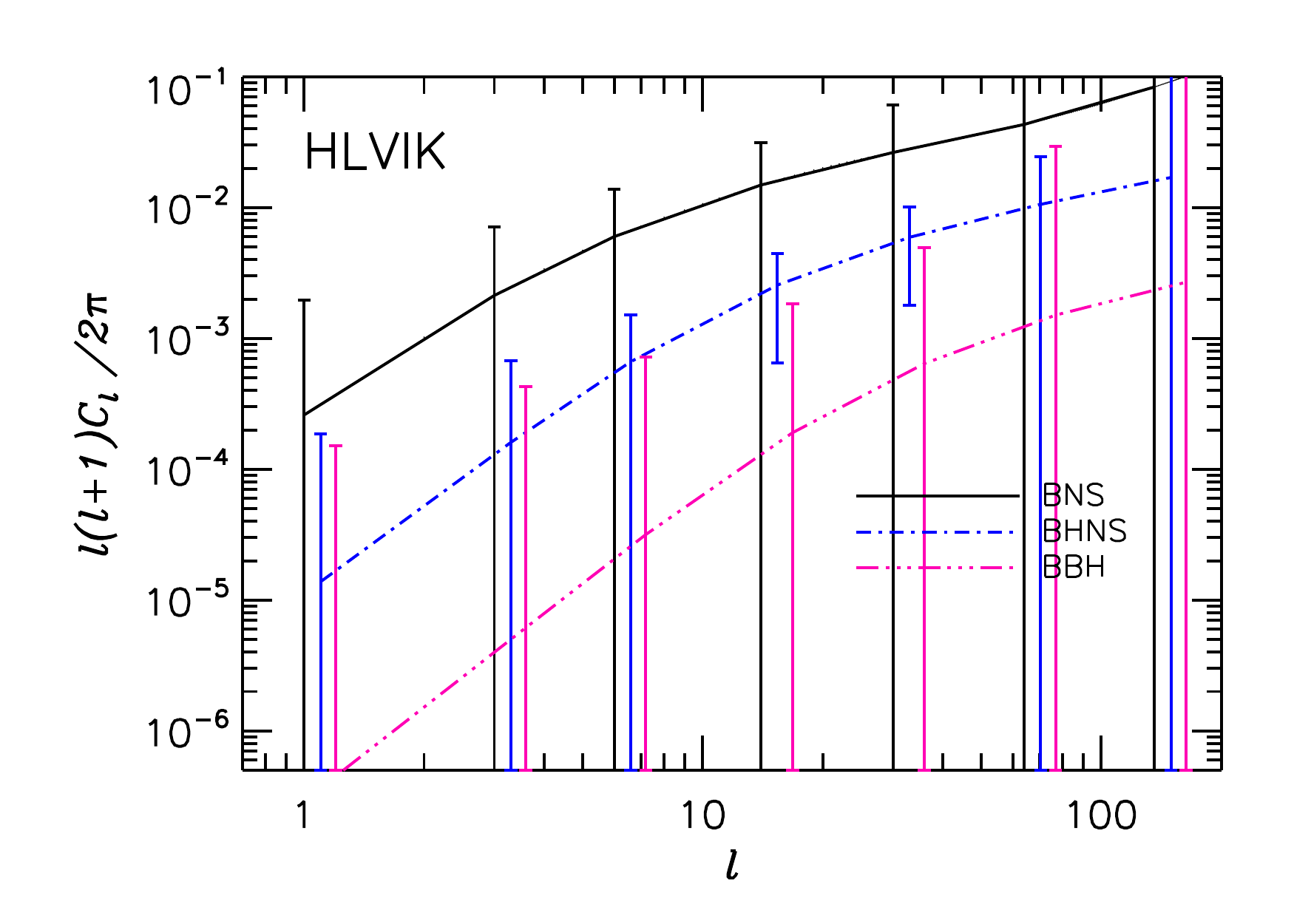} 
\end{tabular}
\end{center}
\caption{Expected autocorrelation  power spectrum (APS) and errors for the three classes of binary mergers considered and for the HLVIK GW detectors configuration.
Note that the error bars for the BHNS and BBH cases have been slightly shifted in $\ell$ to improve the readability of the plot.
\label{fig:autoc}}
\end{figure}

\section{The galaxy catalogs}\label{catalogs}

\subsection{Current catalogs}

Among the catalogs presently available, we will use in the following the 2MASS photometric redshift catalog (2MPZ) \cite{Bilicki:2013sza}, 
the WISE$\times$SuperCosmos catalog (WIxSC) \cite{Bilicki:2016irk}, and 
SDSS Data Release 12 photometric catalog \cite{Alam:2015mbd,beckSDSS2016}. The catalogs characteristics are described in \cite{Cuoco:2017bpv,Stolzner:2017ged}, which the interested reader can consult for details. Their source redshift distributions, ${\rm d}N/{\rm d}z$, are reported in Fig.~\ref{fig:catal}.
These catalogs are chosen in such a way to have a large sky-coverage
and a large number of galaxies, which are crucial characteristics in order to 
reduce the error on the expected cross-correlation (cf. Eq.~\ref{eq:cross}-\ref{eq:noise}) and thus
optimise the 
chances to detect a cross-correlation with GW events.
We provide below a short summary of the main catalogs' features.

\begin{itemize}

\item 2MPZ is a shallow catalog covering the redshift range $z\sim0.0-0.2$, including $\sim$ $8 \times 10^5$ galaxies
and having a large sky-coverage of about 70\%. Given the very large overlap with the BNS and BHNS distributions for the  HLV and HLVIK detector configurations, it 
is extremely suitable for studying their cross-correlation with LSS.

\item WIxSC is an extension of 2MPZ to larger redshifts, up $z\sim 0.4$, with a similar sky-coverage of 70\%, and includes 
about $2\times 10^7$ galaxies. Similar to 2MPZ, it is especially adequate to perform  cross-correlation studies  for  BNS and BHNS classes in the HLV and HLVIK cases.

\item SDSS  has a similar number of galaxies as WIxSC, but with a much smaller sky-coverage (about 20\%), while extending up
to $z\sim0.7$. The SDSS ${\rm d}N/{\rm d}z$ shown in Fig.~\ref{fig:catal} resembles more closely the BBH class of events for the HLV and HLVIK cases (see Fig. \ref{fig:zdistr}),
and is thus better suited for their cross-correlation analysis.

\end{itemize}

\subsection{Future catalogs}

Current surveys are almost all-sky only up $z\sim 0.4$, and with a large statistics and sky-coverage 
only up to $z\sim0.7$ (SDSS). 3G GW observatories will be all-sky and cover up to $z\sim 5$, hence current catalogs would only allow for adequate cross-correlation studies of a small fraction of the collected GW events. This gap is however expected to be filled in the next decade by new surveys like EUCLID, LSST and, to some extent, by SPHEREx. 

EUCLID~\cite{Amendola:2016saw} will have both spectroscopic and photometric redshift samples. 
For the purpose of performing cross-correlations a very precise redshift estimate is not crucial and thus
we consider here only the photometric sample. The advantage is that the number of galaxies 
will be extremely large, about  $\sim 1.6 \times 10^{9}$, with a galaxy density of about 30 arcmin$^{-2}$,
providing a greatly reduced galaxy shot noise, and thus considerably improving the cross-correlation sensitivity.
Further specifics of the EUCLID photometric sample reported in the following are taken from \cite{Amendola:2016saw}. 
Its ${\rm d}N/{\rm d}z$ can be approximated as 
\begin{equation}
   \frac{{\rm d}N}{{\rm d}z} \propto  z^2 \exp \left\{   -    \left(   \frac{z}{z_m/1.412}       \right)^{1.5}    \right\} ,
\end{equation}
with $z_m = 0.9$ the median redshift. As shown in Fig.~\ref{fig:catal},  it has a very good overlap with the redshift distribution of GW events for the 3G detectors benchmark.  The survey area will be $\sim 1.5 \times 10^4$ deg$^2$, so that $f_{\rm fov}\sim0.4$.
The photometric redshifts of the galaxies will be known with an error of $\Delta z \sim 0.05 \, (1+z)$, thus around 10\% at $z=1$. Note that a reasonable approximation for their bias is $b(z)=\sqrt{1+z}$.

The photometric sample of LSST~\cite{Abell:2009aa,Ivezic:2008fe} will have very similar numbers as EUCLID, hence we will not adopt a different galaxy catalog benchmark. The main difference is that LSST,  providing a new survey of the sky every few nights, is optimized to detect transients. Although not directly relevant for the present analysis, this will likely play a crucial role  in identifying the counterparts of the GW events.

Finally, SPHEREx \cite{Dore:2014cca,Dore:2018kgp} will have a similar number of galaxies as EUCLID, with however a larger error  on z, up to $\Delta z \sim 0.2 \, (1+z)$,
although with the possibility to select subsamples at low $z\lesssim 0.5$ with a few \% error.  
Furthermore, it will have almost all-sky coverage, which will be helpful to improve the cross-correlation sensitivity at low $z$ with respect to EUCLID performances.

 \begin{figure*}[th]
\begin{center}
\begin{tabular}{ccc}
 \includegraphics[angle=0,width=0.48\textwidth]{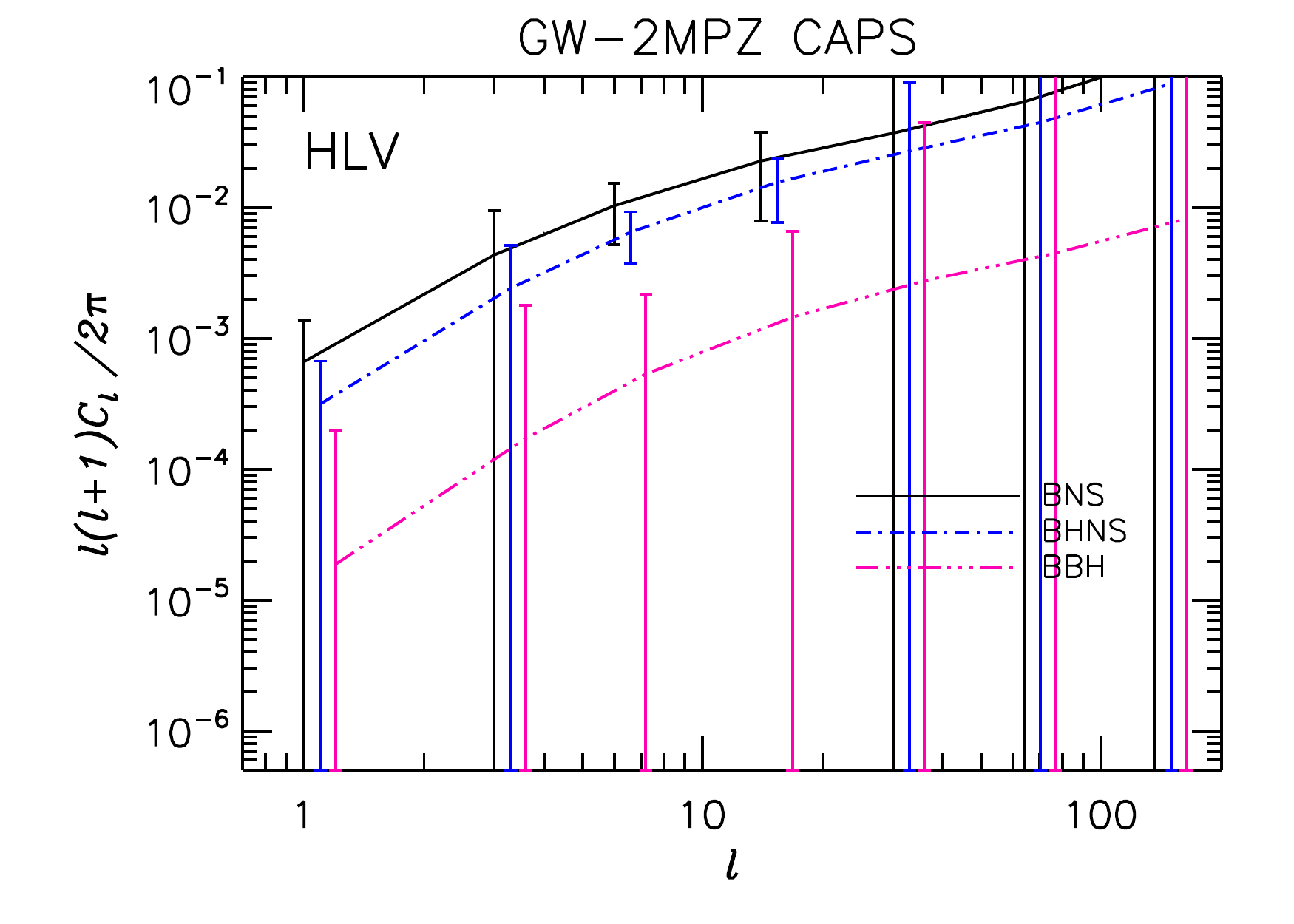}&
 \includegraphics[angle=0,width=0.48\textwidth]{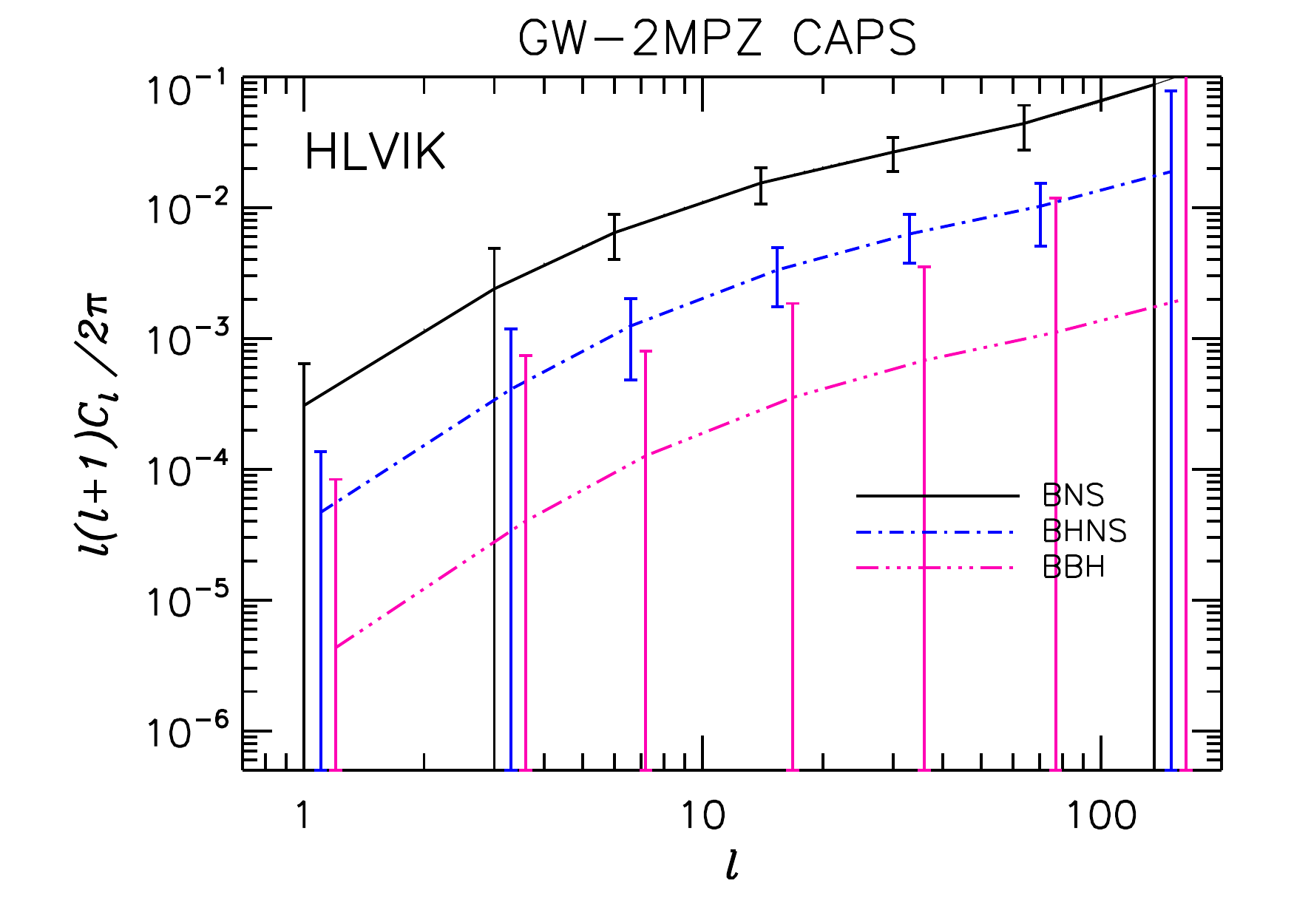}
\end{tabular}
\end{center}
\caption{Expected CAPS and errors between the 2MPZ catalog and 
the three classes of binary mergers considered, for the HLV (left) and HLVIK  (right) GW detectors configurations.
Note that the error bars for the BHNS and BBH cases have been slightly shifted in $\ell$ to improve the readability of the plot.
\label{fig:xcorrel}}
\end{figure*}

 \begin{figure*}[th]
\begin{center}
\begin{tabular}{ccc}
\includegraphics[angle=0,width=0.48\textwidth]{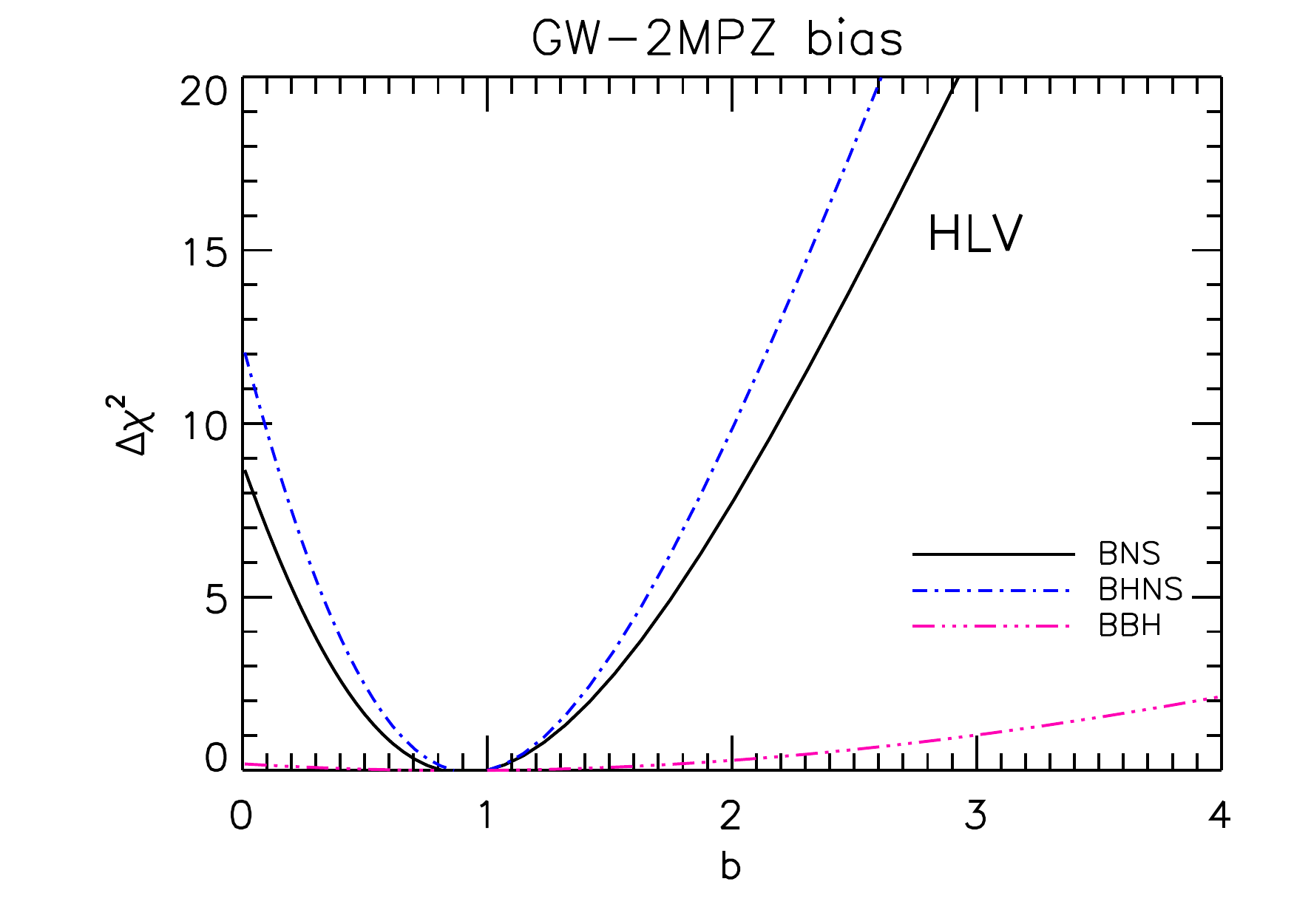} & 
\includegraphics[angle=0,width=0.48\textwidth]{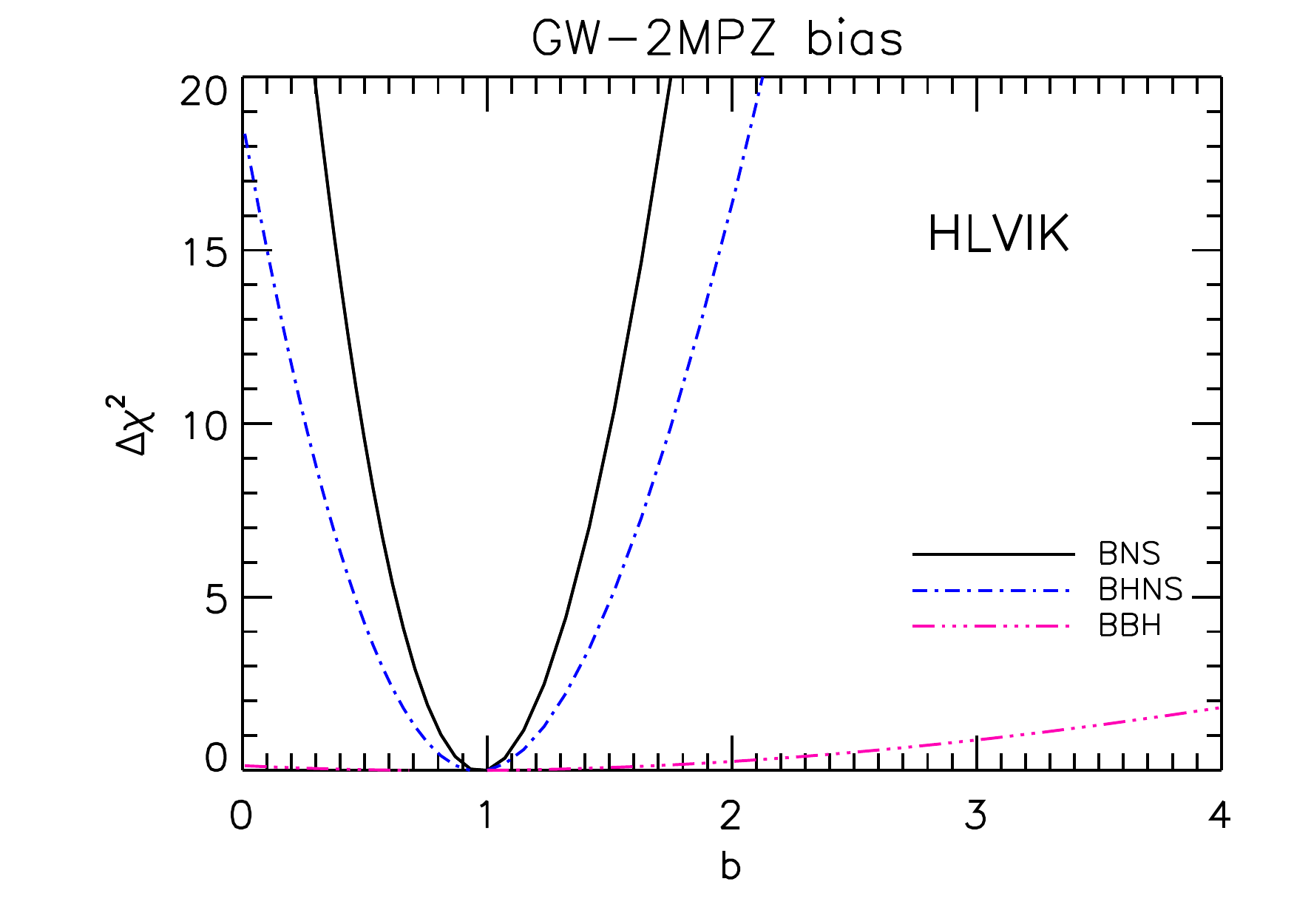}  
\end{tabular}
\end{center}
\caption{$\Delta \chi^2$ values as a function of bias $b$ for the cross-correlation between GW events and the 2MPZ catalog, for the HLV (left) and HVLIK case (right). \label{fig:TS}}
\end{figure*}

\section{Results}\label{results}
\subsection{Auto-correlation of  GW events}

We first briefly mention  the perspectives for measuring  the autocorrelation of the GW events. The results for the HLVIK case are reported in Fig.~\ref{fig:autoc}. 
The plot shows the expected APS and error bars for the three classes of GW coalescence events.
One can see that the normalization of the APS is  larger in the BNS case, which is 
the source population with redshift distribution skewed towards the lowest $z$. This effect is purely geometrical,
and depends on the cosmological volume involved. 
In the limited volume of our local universe, relevant for the BNS case,
the total anisotropy is very large, producing a large APS, while in the cases of
 BBH and BHNS, which extend to larger redshifts,  the averaging over a larger volume reduces the final anisotropy and thus the APS normalization.
Clearly, since a large anisotropy is easier to detect, this 
qualitatively favors the detection of BNS compared to BHNS, and BNS or BHNS compared to BBH.
Needless to say, however, the total number of events is also important in assessing the error with which a measurement can be performed. 
Overall, we find that for 2G GW detectors there is little hope to measure a significant auto-correlation. 
For 3G detectors, instead, we find that 
the perspectives for  detection are a bit brighter.
In this case, the improved statistics may be enough to beat the shot noise, despite the fact that the normalization of the  APS is very low as a result of the redshift distribution being peaked at $z\sim1$ for all type of binaries. 

\subsection{Cross-correlation of 2G GW events}

A much more promising diagnostic for the not-so-far future (i.e. for HLV and especially HLVIK) is the cross-correlation.
The expected cross-correlation with 2MPZ is shown in Fig.~\ref{fig:xcorrel}, and, indeed,
it can be seen that the expected error bars on the CAPS are much smaller for the APS case,
indicating that a significant detection of the cross-correlation signal should be possible.
The reason for this marked improvement can be in part understood looking at
 the $z$ distribution of the expected events (Fig.~\ref{fig:zdistr}) and  the one of the 2MPZ catalog (Fig.~\ref{fig:catal}). 
The overlap between the two is very large at relatively low $z$, especially for BNS and BHNS, 
 thus optimizing the expected cross-correlation.
 Furthermore, the large number of galaxies in the 2MPZ catalog gives a very small shot noise, contributing to 
 reduce the size of the CAPS error bars. 
 For the WIxSC HLV case (not shown), instead, the overlap of the $z$ distributions is less pronounced and   only a marginal detection for the BHNS case seems possible, while the HLVIK case shows a strong cross-correlation signal both with BNS and BHNS, but still a weak one with BBH.
Finally, detection of a cross-correlation with SDSS (not shown) appears out of reach even for HLVIK, where only a BHNS signal may be detected for favorable statistics. These differences highlight the importance of the superposition of the redshift region covered by the galaxy catalogs with the one of the GW sample.

The detectability of the signal can be quantified through the use  of $\chi^2$ as defined in Eq.~(\ref{chi2}).
In particular, in Fig.~\ref{fig:TS} we show the quantity $\Delta\chi^2(b)= \chi^2(b)-\chi^2(b=1)$
as function of the bias $b$, for all type of binaries, 
for HLV  and HLVIK cases, and for the 2MPZ catalog. 
We remind the reader that our reference  model assumes that GW events are an unbiased tracer ($b=1$) of LSS and,
thus,  a model with a given $b$ is compared to the case $b=1$ via the above $\Delta \chi^2$. 
The plot encodes a double information:
\begin{itemize}
\item $\Delta\chi^2(b)$ can be used to make a forecast on the inference on the $b$ parameter. 
The range of $b$ around $b=1$ characterized by a deviation $\Delta\chi^2$=1 is thus the expected (one sigma) error on $b$.
\item  Since  $b=0$ (which represents isotropy of the GW events, and thus no cross-correlation with LSS) is a
nested case with respect to the model with $b$ as free parameter,  Wilks theorem~\cite{Wilks:1938dza} applies and the
quantity $\Delta\chi^2(b=0)$ is expected to behave as a $\chi^2$ distribution with one degree of freedom.
Thus, $\sqrt{\Delta \chi^2(0)}$ quantifies the statistical significance, in $\sigma$'s,
of the cross-correlation with respect to the isotropic sky. 
\end{itemize}
A table with the $\Delta \chi^2(0)$ values for the different combinations of catalogs and GW detector configurations is given in Tab.~\ref{tab:TS}.
Considering  the HLVIK configuration, the cross-correlation of BNS events with the 2MPZ catalog should be detectable with a significance of about $\sqrt{41}\sim 6.5 \sigma$. This catalog is indeed optimal since peaking at low $z$. On the other hand, the  BHNS class would emerge as the one with highest significance (about $9.1 \sigma$) when cross-correlating with WIxSC. The SDSS catalog performs significantly worse, due to the inadequate matching of the $z$ distribution of events. For HLV, the 2MPZ catalog offers a concrete hope for a $\sim 3\sigma$ detection of BNS (and perhaps BHNS at a slightly larger significance) cross-correlation within a decade of running at design sensitivity.
The BBH case, however, remains out of reach, which is mainly the result of a redshift distribution peaking at quite large $z$ where the anisotropy is expected to be small. In this case,  a large number of GW events would be required to be reach a significant cross-correlation detection. However, this can be in part overcome using redshift tomography, as we discuss in the next section.

\begin{table}[t]
\caption{$\Delta \chi^2(0)$ values for the cross-correlation with current galaxy catalogs.}
\label{tab:TS}
\centering
{
\footnotesize
\begin{tabular}{c|ccc|ccc}
\hline
Case      &   &  HLV &      &  &  HLVIK &  \\
\hline
Type      &     BNS & BBH & BHNS       &  BNS & BBH & BHNS   \\
\hline \hline
2MPZ     &     9            &   0.2            &   12            &   41            &   0.1            &   18 \\
WIxSC   &        0.7            &  0.7            &   4            &       17            &   1.8            &   83  \\
SDSS    &       0.0           &    0.4           &   1.0            &     2.4            &   1.4            &   19 \\
\hline
\end{tabular}
}
\end{table}

\begin{table}[t]
\caption{$\Delta \chi^2(0)$ values for the cross-correlation of BBH with a future EUCLID-like catalog.}
\label{tab:TSzbin}
\centering
{
\footnotesize
\begin{tabular}{c|c|c|c}
\hline
Case      &     HLV &     HLVIK  &  ET2CE  \\
\hline
\hline 
0zbin                  &     0.2            &     2.6            &              \\
2zbin                  &        0.9            &  19            &                \\
3zbin                  &       1.8           &    30            &                \\
5zbin                  &       3.6           &    51           &           \\
\hline
9zbin                  &                      &                    &   497        \\
4zbin ($z>1$)    &                      &                    &   101       \\
\hline
\end{tabular}
}
\end{table}

\subsection{Future tomographic studies and 3G perspectives}
In order to overcome the above-mentioned limitations, we move to study the potential of a future EUCLID-like catalog for the only type of binaries for which it would be definitely needed, namely BBH. An important ingredient that we consider is redshift tomography. To evaluate the impact of a tomographic approach, we consider the following cases of redshift binning: 
\begin{itemize}
\item[I.]  {\bf 0zbin}: Cross-correlation between BBH catalogs and the EUCLID-like catalog, without binning in redshift. 
\item[II.] {\bf 2zbin}: Cross-correlation between BBH catalogs and the EUCLID-like catalog where both event samples are split in two redshift bins, $z= 0.0-0.3$ and $z=0.3-1.0$. 
\item[III] {\bf 3zbin}: Same as above, but with a splitting in 3 bins $(z= 0.0-0.2,  0.2-0.4, 0.4-1.0)$.
\item[IV.]  {\bf 5zbin}: Same as above, but with a splitting in 5 bins $(z= 0.0-0.1,  0.1-0.2, 0.2-0.3, 0.3-0.5, 0.5-1.0)$. 
\item[V.] {\bf 9zbin}: Same as above, but with a splitting in 9 bins:  the same 5 bins of case  {\bf 5zbin} plus
 four additional bins $(z= 1.0-1.5,  1.5-2.0, 2.0-3.0, 3.0-5.0)$.
\end{itemize}
In all cases the binning in $z$ was chosen so to roughly provide equal numbers of BBH in each bin.

\begin{figure*}[th]
\begin{center}
\begin{tabular}{ccc}
\includegraphics[angle=0,width=0.45\textwidth]{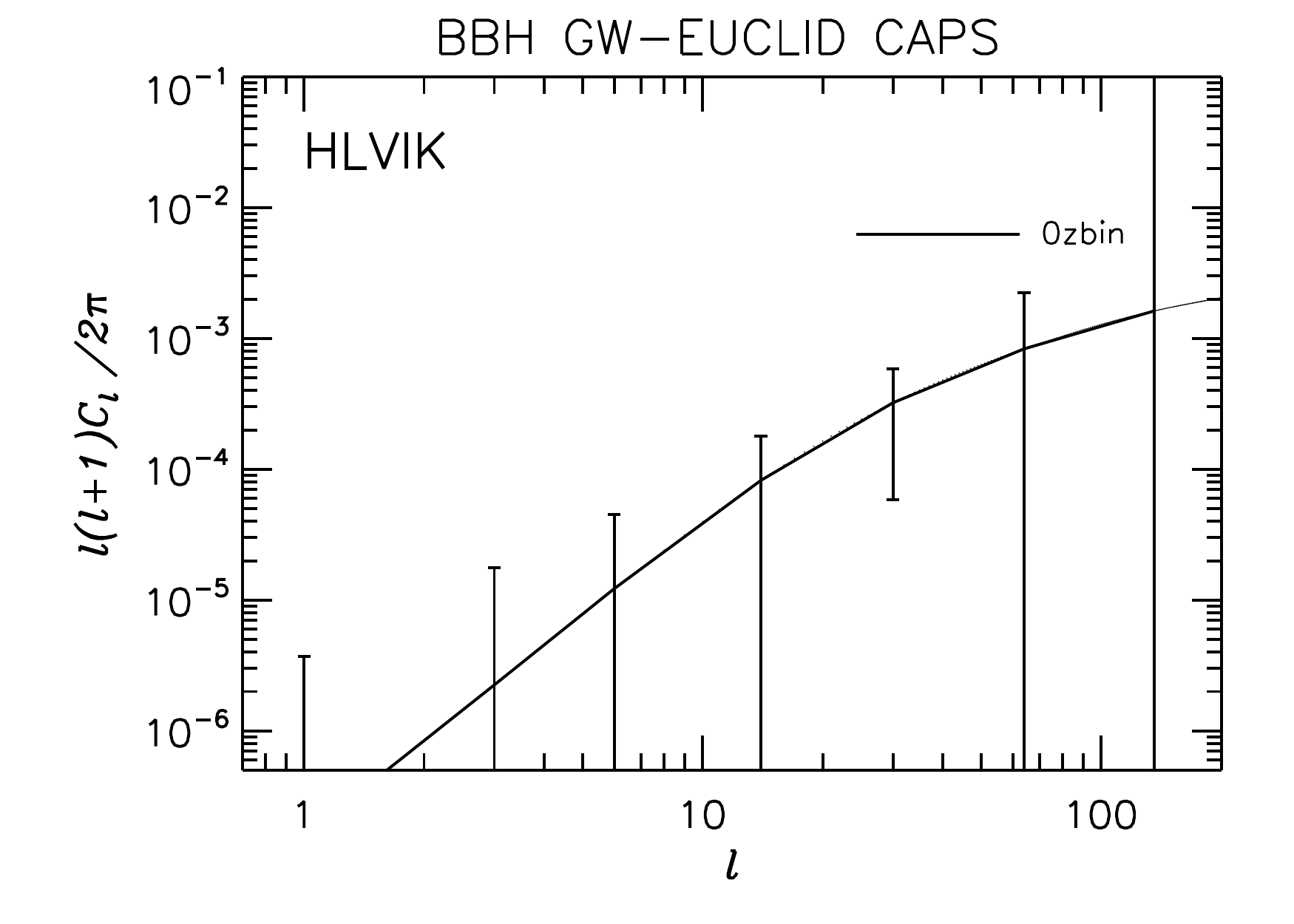} & 
\includegraphics[angle=0,width=0.45\textwidth]{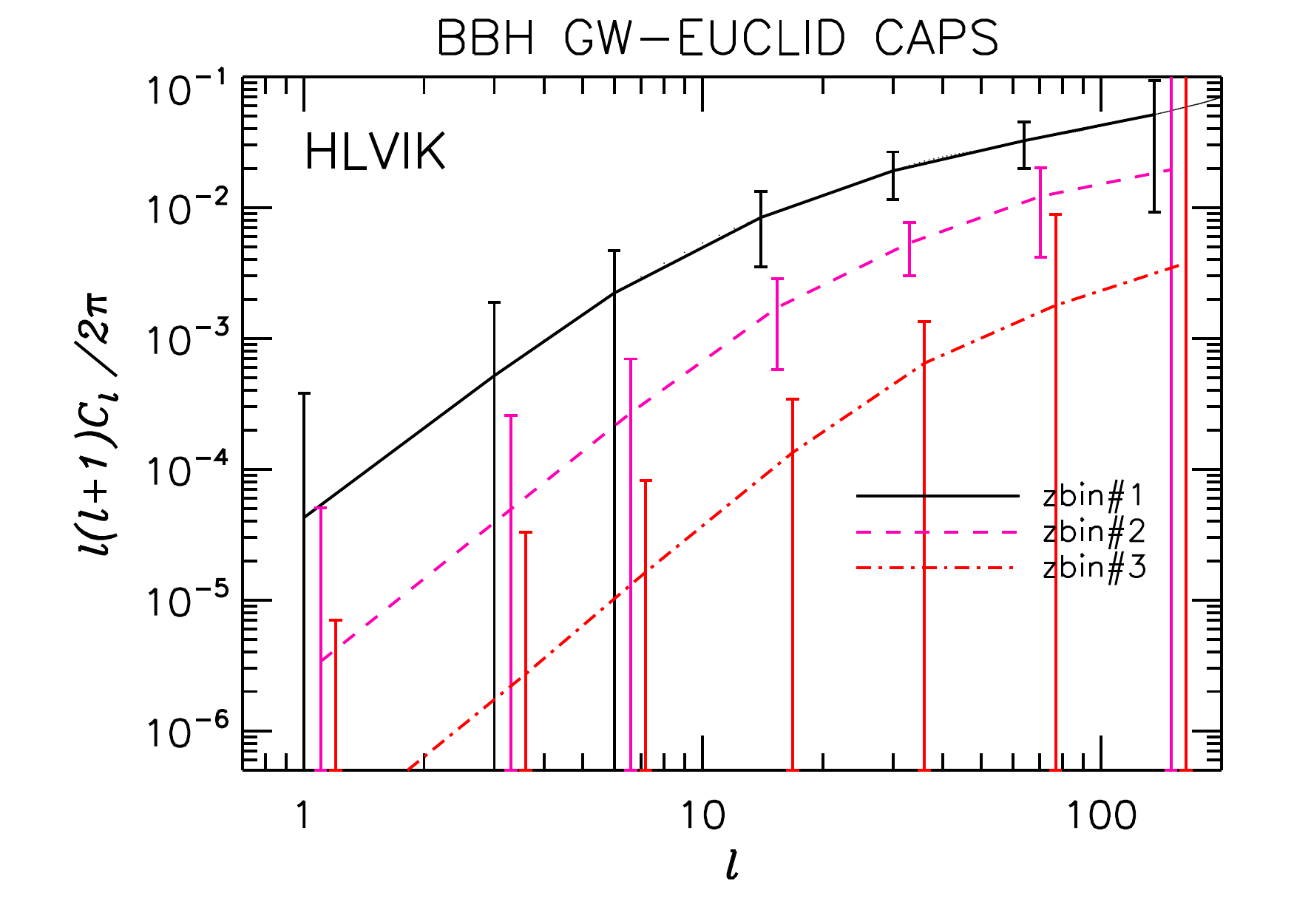} 
\end{tabular}
\end{center}
\caption{Expected cross-correlation between BBH GW events and an EUCLID-like galaxy survey
for the HLVIK case with 10-yr data taking. Left panel: no binning in $z$. Right panel: case
with  3 $z$ bins  with the cross-correlation in each bin  explicitly shown. 
The curves for the different cases have been slightly shifted in $\ell$ to improve the readability of the plot.
Note the overall increase of the signal in the low-$z$ bin(s) and the improved determination of the $C_\ell$ with respect to the left panel. 
\label{fig:EUCL2G}}
\end{figure*}

From Tab.~\ref{tab:TSzbin} we see that for HLV case  
no significant detection of the BBH signal is expected, with a maximum
significance of at most $\sim  1.9 \sigma$ using 5 $z$ bins.
For HLVIK, instead, the impact of EUCLID-like catalogs combined with a tomographic approach can be crucial. Fig.~\ref{fig:EUCL2G} and Tab.~\ref{tab:TSzbin} shows that
with no binning in $z$ only a very modest detection of the signal (at $\sqrt{2.6}\sim 1.6 \sigma$) is possible.
With two bins the significance already significantly increases to $\sim 4.3 \sigma$ and rises to $\sim 7.1 \sigma$
with 5 $z$ bins. We thus see that $z$ binning 
 strongly increases the significance, especially thanks to the bins at low redshift where the anisotropy is the largest. The fact that the CAPS have a larger normalization at low $z$ can be explicitly seen in the right panel of Fig.~\ref{fig:EUCL2G}, where the cross-correlation for  each redshift bin is shown for the case 3zbin.

\begin{figure*}[th]
\begin{center}
\begin{tabular}{ccc}
\includegraphics[angle=0,width=0.48\textwidth]{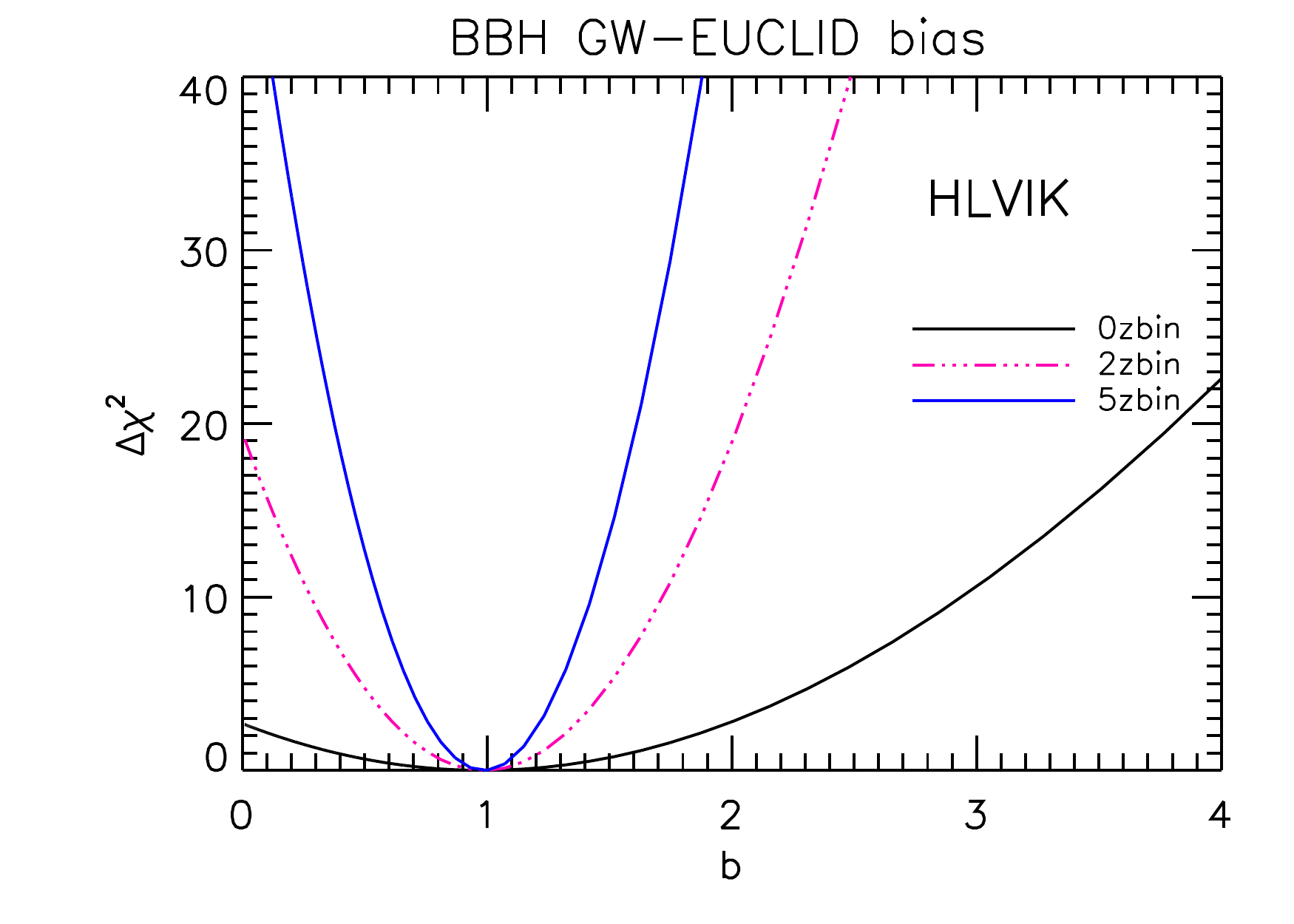} &
\includegraphics[angle=0,width=0.48\textwidth]{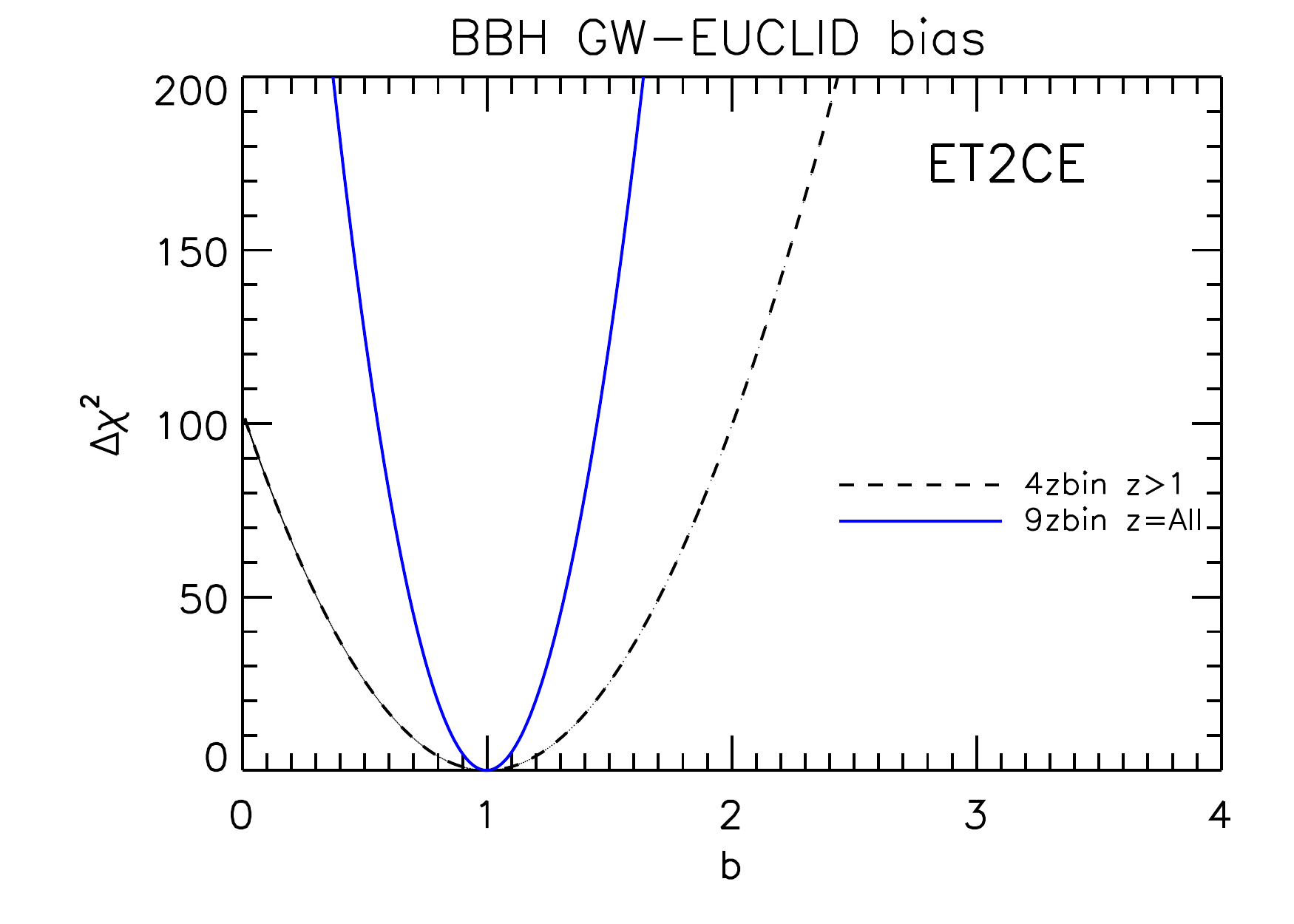} 
\end{tabular}
\end{center}
\caption{$\Delta \chi^2$ as function of $b$ for 10-yr GW data taking for the cross-correlation between BBH GW events and EUCLID-like survey, for HLVIK (left panel) and ET2CE (right panel) cases, and for different binning in $z$. Each panel  illustrates the power of the tomographic approach, with precision  increasing (i.e. increasing curvature of the functions) with more redshift bins.
\label{fig:EUCLTS}}
\end{figure*}

The results for the ET2CE case are shown in the lower part of Tab.~\ref{tab:TSzbin} and in the right panel
of Fig.~\ref{fig:EUCLTS}. It can be seen that the combination of $z$ binning togheter with the large statistics
available with 3G detectors provides a striking significance of 20 $\sigma$ for the BBH-LSS cross-correlation.
With this significance precise studies will be possible as for example studies of the bias as function
of $z$, as discussed more in detail below.
Finally, a further  result is the fact that 
 neglecting the information from $z<1$, as done in~\cite{Scelfo:2018sny} (dashed black curve in the right panel of Fig.~\ref{fig:EUCLTS}), dramatically decreases the significance of the detection (from about 20 $\sigma$  to 10 $\sigma$), even if only $\sim 20\%$ of the events are at $z<1$. A result which further stresses the importance
 of the information at low $z$ where the anisotropy and the cross-correlation is the largest.

 \begin{figure}[thb]
\begin{center}
\begin{tabular}{c}
\includegraphics[angle=0,width=0.45\textwidth]{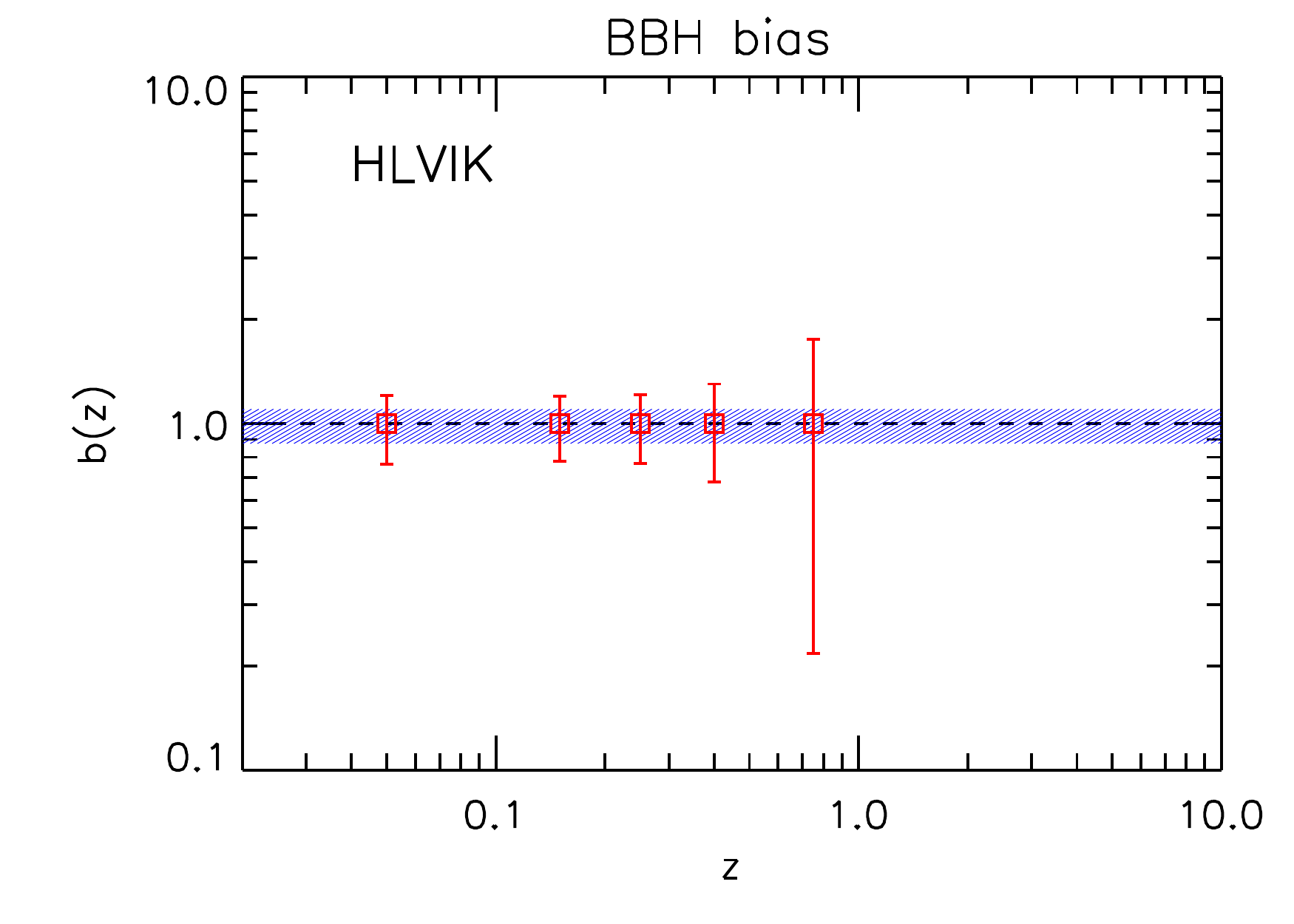} \\
\includegraphics[angle=0,width=0.45\textwidth]{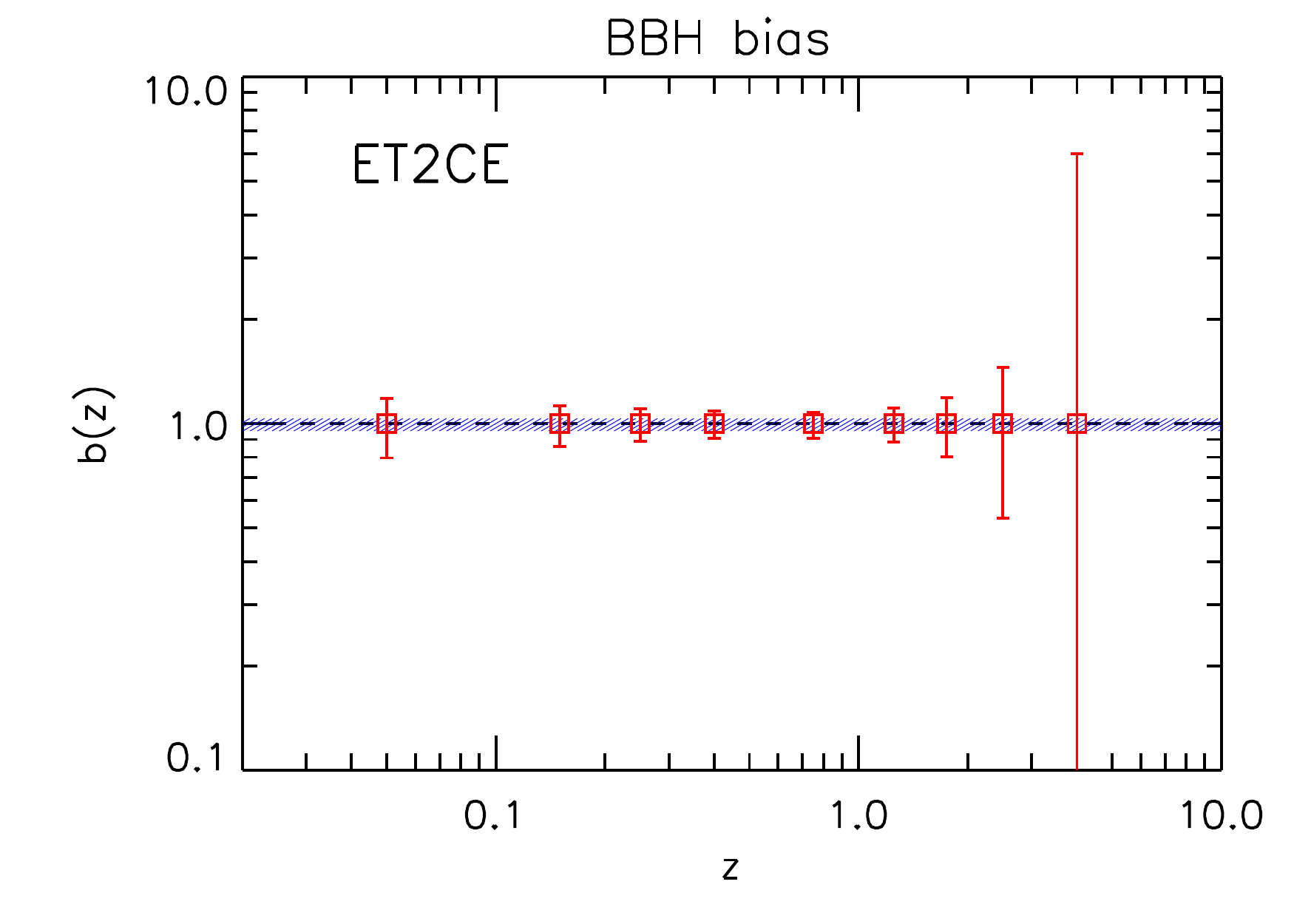} \\
\includegraphics[angle=0,width=0.45\textwidth]{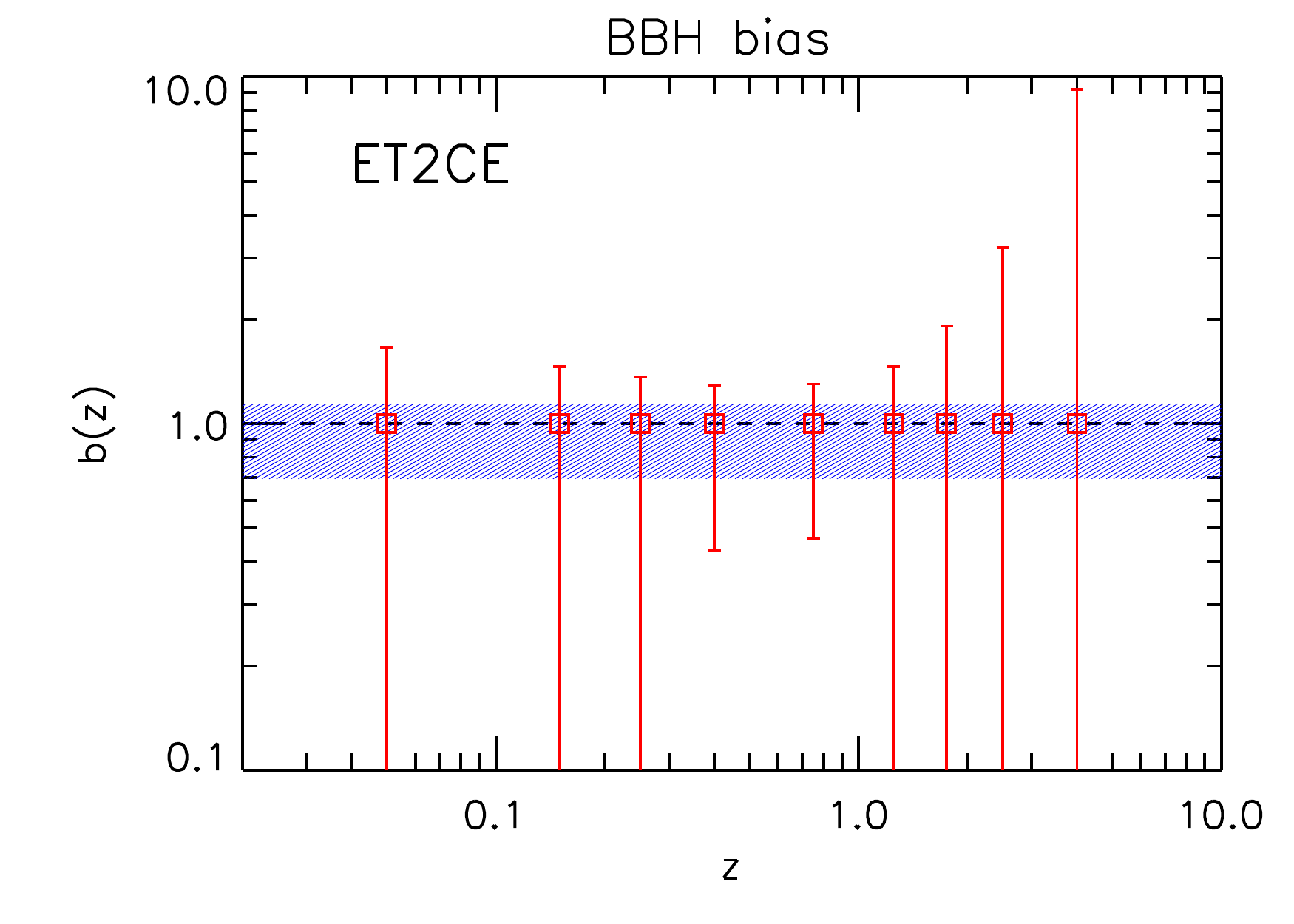} 
\end{tabular}
\end{center}
\caption{Constraint on the BBH bias from the cross-correlation with a EUCLID-like catalog, in each redshift bin separately (error bars) and globally (shaded region), for HLVIK(top panel) and ET2CE (middle). For comparison, the  much less precise bias extracted for ET2CE from the autocorrelation is also shown in the bottom panel. \label{fig:bias}}
\end{figure} 

Fig.~\ref{fig:bias} shows in more detail how well the bias can reconstructed independently in each single bin, for the HLVIK and ET2CE cases.
For the ET2CE case (middle panel) the error is typically $\sim10\%$ at low $z$, decreases to a few percent  in the range around  $z\sim0.5$ and then significantly increases in the range \mbox{$1<z<5$}.
This behavior is the result of two opposite trends: The lower the redshift, the higher the level of anisotropy projected over the sky. On the other hand, the number of events grows with redshift till $z\gtrsim 1$. As a result, there is a sweet spot at intermediate redshifts where the error determination is optimal.
Finally the blue shaded region in Fig.~\ref{fig:bias} shows the combined constraint on the bias considering all the $z$ bins together, yielding an error of a few percent. 
Such a precise BBH bias determination will allow one to eventually discriminate different models at the origin of the BBH population, in particular between an astrophysical or primordial origin. For HLVIK (top panel) the bias will be determined with a precision of about 10\%, definitely enough for a first discrimination among models. In both cases, tomography would allow one to shed light on significant evolutionary effects of the BBH population, if present. 
The lower panel is another illustration of the advantage of the cross-correlation analysis over the auto-correlation one: For the auto-correlation the degradation of the constraints on $b$ for the ET2CE case is
dramatic, and only upper limits on $b$ (or marginal detection for some redshift bins) can be hoped for via this analysis technique.

\section{discussion and conclusions}\label{conclusions}
A few years after the first detection of coalescing binaries via their GW signal, we are about to enter the era of large samples of detected events, allowing one to perform more and more statistical studies. The cross-correlation of the events with LSS (notably galaxy) catalogs has been discussed in the recent past as a potential tool, in particular to identify the underlying population of progenitors of BBH and discriminate, for instance, between astrophysical and primordial models. 

In this article, we have revisited this problem with a number of advances both on the technical and physical side:
i) We have relied on realistic mock GW catalogs, calibrated on the most recent GW data, and on existing (and expected) galaxy catalogs.
ii) We post-processed these events with realistic reconstruction tools for existing (and expected) detectors configurations, in order to determine the statistics as well as the key information concerning the angular resolution. 
iii) We have extended the study to different classes of events, namely BBH, BNS and BHNS.
iv) We have explored the importance of the redshift distribution of both catalogs and GW samples and studied the impact of a tomographic approach.

Our results are rather promising: For BNS, the current generation of detectors (running at design sensitivity for a decade) should be able to marginally detect a cross-correlation with a few hundreds of events, in particular with relatively shallow catalogs like 2MPZ, despite a relatively poor GW localization. When the complete 2G network of five detectors will be available, a firm measurement of the cross-correlation signal will be possible, thanks to both an increased statistics and a much better angular resolution. 
For BBH, however, both current GW detectors and current surveys prove inadequate for a detection of a cross-correlation signal. Fortunately, forthcoming surveys like EUCLID (similar performances are expected for LSST and, possibly, SPHEREx) will provide a much better coverage of the redshift range, where the bulk of BBH events comes from. This, combined with a fully tomographic approach, would allow one to detect the BBH bias at the 10\% level with a network of five 2G detectors, and thus permit the first tests on their origin. Finally, 3G detectors will be able to detect mergers at large $z$, in a range of distances much better covered by forthcoming, deep cosmological surveys.
We foresee that the combination of these two facts will open exciting perspectives for precision studies. 
\vspace{0.2cm}
\begin{acknowledgments}
We thank Walter Del Pozzo for discussions. 
FC, TR, and PDS acknowledge support from IDEX Univ. Grenoble Alpes, under the program {\it Initiatives de Recherche Strat\'egiques},  project ``Multimessenger avenues in gravitational waves'' (PI: PDS).
A.C. is supported by: `Departments of Excellence 2018-2022'' grant awarded by the Italian Ministry of Education, University and Research (MIUR) L. 232/2016; Research grant ``The Dark Universe: A Synergic Multimessenger Approach'' No. 2017X7X85K funded by MIUR; Research grant TAsP (Theoretical Astroparticle Physics) funded by INFN.
\end{acknowledgments}

\bibliography{GWcorrel}

\end{document}